\documentclass[fleqn,usenatbib]{mnras}
\usepackage{newtxtext,newtxmath}

\usepackage[T1]{fontenc}
\usepackage{ae,aecompl}

\usepackage{algorithmic,algorithm}

\usepackage{graphicx}	
\usepackage{amsmath}	
\usepackage{amssymb}	
\usepackage{xcolor}	

\definecolor{dred}{rgb}{0.75,0.,0.}
\definecolor{darkred}{rgb}{0.5,0.,0.}
\definecolor{darkgreen}{rgb}{0.,0.5,0.}

\newcommand{\beq}{\begin{equation}}
\newcommand{\eeq}{\end{equation}}

\newcommand{\pars}{\vec{\theta}}

\newcommand{\mstar}{h^{-1}M_\odot}

\title[ABC in large scale structure]{Approximate Bayesian Computation in Large Scale Structure: constraining the galaxy-halo connection}

\author[Hahn \& Vakili et al.]{
\parbox{\textwidth}{
ChangHoon~Hahn,$^{1,\ast}$\thanks{E-mail: chh327@nyu.edu}
Mohammadjavad~Vakili,$^{1,\ast}$\thanks{E-mail: mjvakili@nyu.edu}
Kilian~Walsh,$^{1}$\\
Andrew~P.~Hearin,$^{2}$
David~W.~Hogg,$^{1,3,4,5}$
and Duncan Campbell,$^{6}$
}
  \vspace*{10pt} \\
$*$ These authors have contributed equally to the paper\\ 
$^{1}$Center for Cosmology and Particle Physics, Department of Physics, New York University, 726 Broadway, New York, NY\\
$^{2}$Yale Center for Astronomy \& Astrophysics, Yale University, New Haven, CT\\
$^{3}$Flatiron institute, 160 Fifth Avenue, New York, NY\\
$^{4}$Center for Data Science, New York University, 60 Fifth Ave, New York, NY 10011\\
$^{5}$Max-Planck-Institut f\"ur Astronomie, K\"onigstuhl 17, D-69117 Heidelberg, Germany\\
$^{6}$Department of Astronomy, Yale University, New Haven, CT
}

\date{Accepted XXX. Received YYY; in original form ZZZ}

\pubyear{2017}

\begin{document}
\label{firstpage}
\pagerange{\pageref{firstpage}--\pageref{lastpage}}
\maketitle

\begin{abstract}

Standard approaches to Bayesian parameter inference in large scale 
structure assume a Gaussian functional form (chi-squared form) 
for the likelihood. This assumption, in detail, cannot 
be correct. Likelihood free inferences such as Approximate Bayesian Computation (ABC) 
relax these restrictions and make inference possible without making any 
assumptions on the likelihood. Instead ABC relies on a forward generative model of the data and a metric 
for measuring the distance between the model and data. In this work, we 
demonstrate that ABC is feasible for LSS parameter inference by using it 
to constrain parameters of the halo occupation distribution (HOD) model 
for populating dark matter halos with galaxies.

Using specific implementation of ABC supplemented with Population Monte Carlo
importance sampling, a generative forward model using HOD, and a distance metric 
based on galaxy number density, two-point correlation function, and galaxy group
multiplicity function, we constrain the HOD parameters of mock observation 
generated from selected ``true'' HOD parameters. The parameter constraints we 
obtain from ABC are consistent with the ``true'' HOD parameters, demonstrating that ABC can be reliably  used for parameter inference in LSS. Furthermore, we compare our ABC constraints to constraints we obtain using a pseudo-likelihood function of Gaussian form with MCMC and find consistent HOD parameter constraints. Ultimately our results suggest that ABC can and should be applied in parameter  inference for LSS analyses. 
\end{abstract}




\section{Introduction}

Cosmology was revolutionized in the 1990s with the introduction of likelihoods---%
pro\-ba\-bil\-ities for the data given the theoretical model---%
for combining data from different surveys and performing principled inferences of
the cosmological parameters (\citealt{White:1996aa, Riess:1998aa}). 
Nowhere has this been more true than in cosmic microwave background (CMB) studies,
where it is nearly possible to analytically evaluate a likelihood function that
involves no (or minimal) approximations (\citealt{Oh:1999aa}, \citealt{Wandelt:2004aa},  
\citealt{Eriksen:2004aa}, \citealt{planckI, planckII}). 

Fundamentally, the tractability of likelihood functions in cosmology flows from
the fact that the initial conditions are exceedingly close to Gaussian in form 
(\citealt{planck_NG, planck_inflation}),
and that many sources of measurement noise are also Gaussian (\citealt{Knox:1995aa, Leach:2008aa}).
Likelihood functions are easier to write down and evaluate when things are closer 
to Gaussian, so at large scales and in the early universe. Hence likelihood analyses 
are ideally suitable for CMB data. 

In large-scale structure (LSS) with galaxies, quasars, and quasar absorption systems as tracers,
formed through nonlinear gravitational evolution and biasing, the likelihood {\em cannot} be Gaussian. 
Even if the initial conditions are perfectly Gaussian, the growth of structure creates non-linearities 
which are non-Gaussian (see \citealt{Bernardeau:2002aa} for a comprehensive review). 
Galaxies form within the density field in some complex manner that is modeled only effectively
(\citealt{Dressler:1980aa, Kaiser:1984aa, Santiago:1992aa, Steidel:1998aa}; see \citealt{somerville15} for a recent review).  
Even if the galaxies were a Poisson sampling of the density field, which they are not (\citealt{Mo:1996aa, Sommerville:2001aa, Casas-Miranda:2002aa}), it would be tremendously difficult to write down even 
an approximate likelihood function (\citealt{devpois}).

The standard approach makes the strong assumption that the likelihood function 
for the data can be approximated by a pseudo-likelihood function that is a Gaussian
probability density in the space of the two-point correlation function estimate. 
It is also typically limited to (density and) two-point correlation 
function (2PCF) measurements, assuming that these measurements constitute 
sufficient statistics for the cosmological parameters. 
As Hogg (in preparation) demonstrates, the assumption of a Gaussian 
pseudo-likelihood function cannot be correct (in detail) at any scale, 
since a correlation function, being related to the 
variance of a continuous field, must satisfy non-trivial positive-definiteness 
requirements. These requirements truncate function space such that the 
likelihood in that function space could never be Gaussian. The failure of this 
assumption becomes more relevant as the correlation function becomes better measured, 
so it is particularly critical on intermediate scales, where neither shot 
noise nor cosmic variance significantly influence the measurement. 

Fortunately, these assumptions are not required for cosmological inferences, 
because high-precision cosmological simulations can be used to directly calculate 
LSS observables. Therefore, we can simulate not just the one- or two-point statistics of the galaxies, but also any higher order statistics that might provide additional constraining power on a model. In principle, there is therefore no strict need to rely on these common but specious analysis  assumptions as it is possible to calculate a likelihood function directly from simulation outputs.

Of course, any naive approach to sufficiently simulating the data would be ruinously
expensive. Fortunately, there are principled, (relatively) efficient methods for 
minimizing computation and delivering correct posterior inferences, using only a 
data simulator and some choices about statistics. 
In the present work, we use Approximate Bayesian Computation---ABC---which provides a \emph{rejection sampling} 
framework (\citealt{abcrejectionsampling}) that relaxes the assumptions of the traditional approach. 

ABC approximates the posterior probability distribution function (model given the data)
by drawing proposals from the prior over the model parameters, simulating the data from the 
proposals using a forward generative model, and then rejecting the proposals that are beyond 
a certain threshold ``distance'' from the data, based on summary statistics of the data. 
In practice, ABC is used in conjunction with a more efficient sampling operation like 
Population Monte Carlo (PMC; \citealt{smc}). 
PMC initially rejects the proposals from the prior with a relatively large ``distance'' threshold. 
In subsequent steps, the threshold is updated adaptively, and samples from the proposals that have 
passed the previous iteration are subjected to the new, more stringent, threshold criterion (\citealt{abcpmc}). 
In principle, the distance metric can be any positive definite function that compares 
various summary statistics between the data and the simulation.  

In the context of astronomy, this approach has been used in a wide range of topics including 
image simulation calibration for wide field surveys (\citealt{abccosmology}),
the study of the morphological properties of galaxies at high redshifts (\citealt{abcmorphology}),
stellar initial mass function modeling (Cisewski et al. in preparation),
and cosmological inference with  
with weak-lensing peak counts (\citealt{abcwl,abcwl2}), Type Ia Supernovae (\citealt{abcsn}), 
and galaxy cluster number counts (\citealt{cosmoabc}). 

In order to demonstrate that ABC can be tractably applied to parameter estimation in contemporary 
LSS analyses, we narrow our focus to inferring the parameters of a Halo Occupation Distribution (HOD) 
model. The foundation of HOD predictions is the halo model of LSS, that is, 
collapsed dark matter halos are biased tracers of the underlying cosmic density field 
(\citealt{press74, bond91, cooray_sheth2002}). The HOD specifies how the dark matter 
halos are populated with galaxies by modeling the probability that a given halo hosts 
$N$ galaxies subject to some observational selection criteria 
(\citealt{lemson99, seljak2000,scoccimarro2001,berlind_weinberg2002,zheng2005}).
This statistical prescription for connecting galaxies to halos has been remarkably
successful in reproducing the galaxy clustering, galaxy--galaxy lensing, and other 
observational statistics (\citealt{Rodriguez-Torres:2015aa, miyatake15}), and is a 
useful framework for constraining cosmological parameters (\citealt{vdb03, tinker05, cacciato13, more13}) 
as well as galaxy evolution models (\citealt{conroy09, Tinker:2011aa, leauthaud12, behroozi13, Tinker:2013aa}, 
Walsh et al. in preparation). 

More specifically, we limit our scope to a likelihood analysis of HOD model parameter space, 
keeping cosmology fixed. We forward model galaxy survey data by populating pre-built dark 
matter halo catalogs obtained from high resolution N-body simulations (\citealt{bolshoi,multidark}) using 
$\mathtt{Halotools}$\footnote{http://halotools.readthedocs.org} 
(\citealt{Hearin:2016aa}), 
an open-source package for modeling the galaxy-halo connection. 
Equipped with the forward model, we use summary statistics such as 
number density, two-point correlation function, galaxy group multiplicity function (GMF)
to infer HOD parameters using ABC.  

In Section \ref{sec:method} we discuss the algorithm of the ABC-PMC prescription we use in our analyses. 
This includes the sampling method itself, the HOD forward model, and the computation of summary statistics. 
Then in Section \ref{sec:mock_obv}, we discuss the mock galaxy catalog, which we treat as observation. 
With the specific choices of ABC-PMC ingredients, which we describe in Section \ref{sec:abcpmc_spec}, 
in Section \ref{sec:abc_results} we present the results of our parameter inference using 
two sets of summary statistics, number density and 2PCF and number density and GMF. 
We also include in our results, analogous parameter constraints from the standard MCMC approach, 
which we compare to ABC results in detail, Section \ref{sec:abcvsmcmc}. Finally, we discuss and 
conclude in Section \ref{sec:discussion}.

\section{Methods}\label{sec:method}
\subsection{Approximate Bayesian Computation} \label{sec:abc}
ABC is based on rejection sampling, so we begin this section with a brief overview of 
rejection sampling. Broadly speaking, rejection sampling is a Monte Carlo method 
used to draw samples from a probability distribution, $f(\alpha)$, which is difficult to directly sample. The strategy is to draw samples from an instrumental distribution $g(\alpha)$ that satisfies the condition $f(\alpha) < M g(\alpha)$ for all $\alpha,$ where $M > 1$ is some scalar multiplier. The purpose of the instrumental distribution $g(\alpha)$ is that it is easier to sample than $f(\alpha)$ (see \citealt{bishop} and refernces therein). 

In the context of simulation-based inference, 
the ultimate goal is to sample from the joint probability of a
simulation $X$ and parameters $\pars$ given observed data $D$, the
posterior probability distribution. From Bayes rule this posterior 
distribution can be written as 
\beq
p(\pars, X | D) = \frac{p(D|X)p(X|\pars)\pi(\pars)}{\mathcal{Z}}
\eeq
where $\pi(\pars)$ is the prior distribution over the parameters of 
interest and $\mathcal{Z}$ is the evidence, 
\beq
\mathcal{Z} = \int d\pars \; dX\; p(D|X) p(X|\pars) \pi(\pars), 
\eeq
where the domain of the integral is all possible values of $X$ and $\pars$. 
Since $p(\pars, X | D)$ cannot be directly sampled, we use rejection 
sampling with instrumental distribution 
\beq
q(\pars, X) = p(X|\pars) \pi(\pars)
\eeq
and the choice of 
\beq
M = \frac{\mathrm{max}\; p(D|X)}{\mathcal{Z}} > 1.
\eeq
Note that we do not ever need to know $\mathcal{Z}$. 
The choices of $q(\pars, X)$ and $M$ satisfy the condition 
\beq
p(\pars, X | D) < M q(\pars, X)
\eeq
so we can sample $p(\pars, X | D)$ by drawing ${\pars, X}$ from $q(\pars, X)$.
In practice, this is done 
by first drawing $\pars$ from the prior $\pi(\pars)$ and then generating a 
simulation $X = f(\pars)$ via the forward model. Then ${\pars, X}$ 
is accepted if
\beq \label{eq:reject_samp}
\frac{p(\pars, X | D)}{M q(\pars, X)} = \frac{p(D|X)}{\mathrm{max}\;p(D|X)} > u 
\eeq
where $u$ is drawn from $\mathtt{Uniform}[0,1]$. By repeating this rejection sampling process, 
we sample the distribution $p(\pars, X |D)$ with the set of $\pars$ and $X$ that are accepted. 

At this stage, ABC distinguishes itself by postulating that $p(D|X)$, 
the probability of observing data $D$ given simulation $X$ 
({\em not} the likelihood), is proportional to the 
probability of the distance between the data and the simulation X being less than 
an arbitrarily small threshold $\epsilon$ 
\beq
p(D|X) \propto p(\rho(D,X)<\epsilon)
\label{eq:abc_condition}
\eeq
where $\rho(D, X)$ is the distance between the data $D$ and simulation $X$. 
Eq. \ref{eq:abc_condition} along with the rejection sampling acceptance criteria 
(Eq. \ref{eq:reject_samp}), leads to the acceptance criteria for ABC: $\pars$ is accepted if $\rho(D, X) < \epsilon$. 

The distance function is a positive definite function that measures the closeness 
of the data and the simulation. The distance can be a vector with multiple components 
where each component is a distance between a single summary statistic of the data 
and that of the simulation. In that case, the threshold $\epsilon$ in 
Eq. \ref{eq:abc_condition} will also be a vector with the same dimensions.  
$\pars$ is accepted if the distance vector is less than the threshold vector for 
every component.

The ABC procedure begins, in the same fashion as rejection sampling, by drawing 
$\pars$ from the prior distribution $\pi(\pars)$. The simulation is generated from 
$\pars$ using the forward model, $X = f(\pars)$. Then the distance between 
the data and simulation, $\vec\rho(D, X)$, is calculated and compared to 
$\vec\epsilon$. If $\vec\rho(D, X) < \vec\epsilon$, $\pars$ is accepted. 
This process is repeated until we are left with a sample of $\pars$ that all 
satisfy the distance criteria. This final ensemble approximates the posterior 
probability distribution $p(\pars, X|D)$. 

As it is stated, the ABC method poses some practical challenges. If the 
threshold $\epsilon$ is arbitrarily large, the algorithm essentially 
samples from the prior $\pi(\pars)$. Therefore a sufficiently small threshold
is necessary to sample from the posterior probability distribution. However,
an appropriate value for the threshold is not known \emph{a priori}. Yet, 
even if an appropriate threshold is selected, a small threshold requires 
the entire process to be repeated for many draws of $\pars$ from $\pi(\pars)$ 
until a sufficient sample is acquired. This often presents computation challenges. 

We overcome some of the challenges posed by the above ABC method
by using a Population Monte Carlo (PMC) algorithm as our sampling technique. 
PMC is an iterative method that performs rejection sampling over a 
sequence of $\pars$ distributions ($\{p_1(\pars), ..., p_T(\pars)\}$ for 
$T$ iterations), with a distance threshold that decreases at each iteration of 
the sequence. 

\begin{algorithm} 
\caption{The procedure for ABC-PMC}
\begin{algorithmic}[1] \label{alg:abcpmc}
\IF{$t=1:$}
\FOR{$i=1,...,N$}
   \STATE // \emph{This loop can now be done in parallel for all i}
   \WHILE{$\rho(X,D)>\epsilon_t$}
   \STATE $\pars^{*}_{t} \gets \pi(\pars)$
   \STATE $X = f(\pars^{*}_{t})$
   \ENDWHILE
   \STATE $\pars^{(i)}_{t} \gets \pars^{*}_{t}$
   \STATE $w^{(i)}_{t} \gets 1/N$
\ENDFOR
\ENDIF
\IF{$t=2,...,T:$}
\FOR{$i=1,...,N$}
   \STATE // \emph{This loop can now be done in parallel for all i}
   \WHILE{$\rho(X,D)>\epsilon_t$}
   \STATE Draw $\pars^{*}_{t}$ from $\{\pars_{t-1}\}$ with probabilities $\{w_{t-1}\}$
   \STATE $\pars^{*}_{t} \gets K(\pars^{*}_{t},.)$
   \STATE $X = f(\pars^{*}_{t})$
   \ENDWHILE
   \STATE $\pars^{(i)}_{t} \gets \pars^{*}_{t}$
   \STATE $w^{(i)}_{t} \gets \pi(\pars^{(i)}_{t}) / \big(\sum\limits_{j=1}^{N}w_{t-1}^{(i)}K(\pars^{(j)}_{t-1},\pars^{(i)}_{t}) \big)$
\ENDFOR
\ENDIF
\end{algorithmic}
\end{algorithm}

As illustrated in Algorithm \ref{alg:abcpmc}, for the first iteration $t = 1$, 
we begin with an arbitrarily large distance threshold $\epsilon_1$. We 
draw $\pars$ (hereafter referred to as particles) from the prior distribution 
$\pi(\pars)$. We forward model the simulation $X = f(\pars)$, calculate the 
distance $\rho(D, X)$, compare this distance to $\epsilon_1$, and then 
accept or reject the $\pars$ draw. Because we set $\epsilon_1$ arbitrarily large, 
the particles essentially sample the prior distribution. This process 
is repeated until we accept $N$ particles. We then assign equal weights to 
the $N$ particles: $w_1^i = 1/N$.

For subsequent iterations ($t > 1$) the distance threshold is set such that
$\epsilon_{i,t} < \epsilon_{i,t-1}$ for all components $i$. Although there is 
no general prescription, the distance threshold $\epsilon_{i,t}$ can be 
assigned based on the empirical distribution of the accepted distances of the 
previous iteration, $t-1$. In \citealt{abcsn}, for instance, the threshold of 
the second iteration is set to the $25^\mathrm{th}$ percentile of the distances 
in the first iterations; afterwards in the subsequent iterations, $t$, $\epsilon_{t}$ 
is set to the $50^\mathrm{th}$ percentile of the distances in the previous $t-1$ iteration. 
Alternatively, \citealt{abcwl} set $\epsilon_{t}$ to the median of the distances from 
the previous iteration. In Section \ref{sec:abcatwork}, we describe our prescription 
for the distance threshold, which follows \citealt{abcwl}. 

Once $\epsilon_t$ is set, we draw a particle from the previous 
weighted set of particles ${\pars}_{t-1}$. 
This particle is perturbed by a kernel, set to the covariance of ${\pars}_{t-1}$.
Then once again, we generate a simulation by forward modeling $X = f(\pars^i)$, 
calculate the distance $\rho(X, D)$, and compare the distance to the new distance 
threshold ($\epsilon_t$) in order to accept or reject the particle. This process is 
repeated until we assemble a new set of $N$ particles ${\pars}_t$. We then update the 
particle weights according to the kernel, the prior distribution, and the 
previous set of weights, as described in Algorithm \ref{alg:abcpmc}. The 
entire procedure is then repeated for the next iteration, $t+1$.

There are a number of ways to specify the perturbation kernel in the ABC-PMC algorithm. 
A widely used technique is to define the perturbation kernel as a multivariate Gaussian 
centered on the weighted mean of the particle population with a covariance matrix set to 
the covariance of the particle population. This perturbation kernel is often called the 
global multivariate Gaussian kernel. For a thorough discussion of various schemes for 
specifying the perturbation kernel, we refer the reader to \citealt{optimalkernel}. 

The iterations continue in the ABC-PMC algorithm until convergence is confirmed. 
One way to ensure convergence is to impose a threshold for the acceptance ratio, 
which is measured in each iteration. The acceptance ratio is the ratio of the number 
of proposals accepted by the distance threshold, to the full number 
of proposed particles at every step. Once the acceptance ratio for 
an iteration falls below the imposed threshold, the algorithm has converged and is
suspended. Another way to ensure convergence is by monitoring the fractional change in 
the distance threshold ($\epsilon_t/\epsilon_{t-1} - 1$)
after each iteration. When the fractional change becomes smaller than some 
specified tolerance level, the algorithm has reached convergence. Another 
convergence criteria, is through the derived uncertainties of the inferred
parameters measured after each iteration. When the uncertainties stabilize 
and show negligible variations, convergence is ensured. In Section \ref{sec:abcpmc_spec} 
we detail the specific convergence criteria used in our analysis. 

\subsection{Forward model}\label{sec:forwardmodel}
\subsubsection{Halo Occupation Modeling}

\newcommand{\lcdm}{\Lambda {\rm CDM}}
\newcommand{\dd}{\mathrm{d}}
\newcommand{\mean}[2]{\left\langle#1 \vert {#2}\right\rangle}

\newcommand{\ngal}{N_{\mathrm{g}}}
\newcommand{\nsat}{N_\mathrm{s}}
\newcommand{\ncen}{N_\mathrm{c}}
\newcommand{\pnm}[2]{p(#1|#2)}

\newcommand{\mhalo}{M_{\rm h}}
\newcommand{\mvir}{M_\mathrm{vir}} 

\newcommand{\dndmvir}{\frac{\dd n}{\dd\mvir}}
\newcommand{\dndmhalo}{\frac{\dd n}{\dd\mhalo}}
\newcommand{\dndmvirprime}{\frac{\dd n}{\dd\mvir'}}

\newcommand{\xigg}{\xi_{\mathrm{gg}}}
\newcommand{\xihh}{\xi_{\mathrm{hh}}}
\newcommand{\xiggr}{\xi_{\mathrm{gg}}(r)}
\newcommand{\xiggroneh}{\xi^{1h}_{\mathrm{gg}}(r)}
\newcommand{\xiggrtwoh}{\xi^{2h}_{\mathrm{gg}}(r)}
\newcommand{\ngalaxy}{\bar{n}_{\mathrm{g}}}
\newcommand{\gmf}{\mathcal{\zeta}_{\rm g}}

ABC requires a forward generative model. In large scale structure studies, this implies a model
that is able to generate a galaxy catalog. We then calculate and compare summary statistics of the data and model catalog in an identical fashion
In this section, we describe the forward generative model we use within the framework of the 
halo occupation distribution.

The assumption that galaxies reside in dark matter halos is the bedrock underlying 
all contemporary theoretical predictions for galaxy clustering. The Halo Occupation Distribution 
(HOD) is one of the most widely used approaches to characterizing this galaxy-halo connection. 
The central quantity in the HOD is $\pnm{\ngal}{\mhalo}$, the probability that a halo of mass 
$\mhalo$ hosts $\ngal$ galaxies. 

The most common technical methods for estimating the theoretical galaxy 2PCF utilize the 
first two moments of $P$, which contain the necessary information to calculate the one- 
and two-halo terms of the galaxy correlation function:
\begin{eqnarray}
\label{eq:onehaloterm}
1+\xiggroneh \simeq \frac{1}{4\pi{}r^{2}\ngalaxy^{2}}\int\dd\mhalo\dndmhalo\Xi_{\rm gg}(r|\mhalo) \times \mean{\ngal(\ngal-1)}{\mhalo},
\end{eqnarray} and

\begin{eqnarray}
\label{eq:twohaloterm}
\xiggrtwoh \simeq \xi_{\mathrm{mm}}(r)\left[\frac{1}{\ngalaxy}\int\dd\mhalo\dndmhalo \mean{\ngal}{\mhalo}b_{\mathrm{h}}(\mhalo)\right]^{2}
\end{eqnarray}
In Eqs.~(\ref{eq:onehaloterm}) and (\ref{eq:twohaloterm}), $\ngalaxy$ is the galaxy number density,
$\dd\mathrm{n}/\dd\mhalo$ is the halo mass function, the spatial bias of dark matter halos is 
$b_{\mathrm{h}}(\mhalo),$ and $\xi_{\rm mm}$ is the correlation function of dark matter.  
If we represent the spherically symmetric intra-halo distribution of galaxies by a unit-normalized 
$n_{\rm g}(r),$ then the quantity $\Xi_{\rm gg}(r)$ appearing in the above two equations 
is the convolution of $n_{\rm g}(r)$ with itself. These fitting functions are calibrated using $N$-body 
simulations. 

Fitting function techniques, however, require many simplifying assumptions. For example, 
Eqs.~(\ref{eq:onehaloterm}) and (\ref{eq:twohaloterm}) assume that the galaxy 
distribution within a halo is spherically symmetric. These equations 
also face well-known difficulties of properly treating halo exclusion and scale-dependent bias, 
which results in additional inaccuracies commonly exceeding the $10\%$ level (\citealt{vdBosch13}). 
Direct emulation methods have made significant improvements in precision and accuracy in recent 
years (\citealt{coyote2,coyote1}); however, a labor- and computation-intensive interpolation 
exercise must be carried out each time any alternative statistic is explored, which is one of the goals of the present work.

To address these problems, throughout this paper we make no appeal to fitting functions or emulators. 
Instead, we use the $\mathtt{Halotools}$ package to populate dark matter halos with mock galaxies and 
then calculate our summary statistics directly on the resulting galaxy catalog with the same estimators 
that are used on observational data (\citealt{Hearin:2016aa}). Additionally, through our forward modeling approaching, we are
able to explore observables beyond the 2PCF, such as the group multiplicity function, for which 
there is no available fitting function. This framework allows us to use group multiplicity function for providing quantitative constraints on the galaxy-halo connection. In the following section, we will show that using this observable, we can obtain constraints on the HOD parameters comparable to those found from the 2PCF measurements. 

For the fiducial HOD used throughout this paper, we use the model described in \citealt{zheng07}. 
The occupation statistics of central galaxies follow a nearest-integer distribution with first 
moment given by 
\begin{equation}
\label{eq:ncen}
\langle N_{\mathrm{cen}}\rangle = \frac{1}{2} \left[ 1 + \mathrm{erf}\left(\frac{\log M - \log M_{\mathrm{min}}}{\sigma_{\log M}}\right)\right].
\end{equation}
Satellite occupation is governed by a Poisson distribution with the mean given by 
\begin{equation}
\label{eq:nsat}
\langle N_{\mathrm{sat}}\rangle =  \langle N_{\mathrm{cen}} \rangle \left(\frac{M-M_{0}}{M_1}\right)^{\alpha}.
\end{equation}
We assume that central galaxies are seated at the exact center of the host dark matter halo and 
are at rest with respect to the halo velocity, defined according to $\mathtt{Rockstar}$ halo finder (\cite{rockstar})
as the mean velocity of the inner $10\%$ of particles in the halo. Satellite galaxies are confined to 
reside within the virial radius following an NFW spatial profile (\citealt{nfw}) with a concentration 
parameter given by the $c(M)$ relation (\citealt{nfw_c(M)}). The peculiar velocity of satellites with 
respect to their host halo is calculated according to the solution of the Jeans equation of an NFW 
profile (\citealt{more2010}). We refer the reader to \cite{hearin15}, 
\cite{Hearin:2016aa}, and \url{http://halotools.readthedocs.io}
for further details.  

For the halo catalog of our forward model, we use the publicly available $\mathtt{Rockstar}$ 
(\citealt{rockstar}) halo catalogs of the $\mathtt{MultiDark}$ cosmological $N$-body simulation 
(\citealt{multidark}).\footnote{In particular, we use the {\tt halotools\_alpha\_version2} version of this catalog, made publicly available as part of {\tt Halotools}.} $\mathtt{MultiDark}$ is a collision-less dark-matter only $N$-body simulation. 
The $\Lambda$CDM cosmological parameters of $\mathtt{MultiDark}$ are $\Omega_m = 0.27$, $\Omega_{\Lambda}=0.73$,
$\Omega_{b}=0.042$, $n_{s}=0.95$, $\sigma_{8} = 0.82$, and $h = 0.7$. The gravity solver used in the 
$N$-body simulation is the Adaptive Refinement Tree code (ART; \citealt{art}) run on $2048^3$ particles in a
$1\; h^{-1}\mathrm{Gpc}$ periodic box. $\mathtt{MultiDark}$ particles have a mass of $m_{p} \simeq 8.72 \times 
10^{8} \; h^{-1}M_{\odot}$; the force resolution of the simulation is $\epsilon \simeq 7 h^{-1}$ kpc.

One key detail of our forward generative model is that when we populate the $\mathtt{MultiDark}$ halos 
with galaxies, we do not populate the entire simulation volume. Rather, we divide the volume into a 
grid of $125$ cubic subvolumes, each with side lengths of $200\;h^{-1}\mathrm{Mpc}$. We refer to these 
subvolumes as $\{\mathtt{BOX1}, ..., \mathtt{BOX125}\}$. 
The first subvolume is reserved to generate the mock observations which we describe in Section 
\ref{sec:mock_obv}. When we simulate a galaxy catalog for a given $\pars$ in parameter space, 
we randomly select one of the subvolumes from $\{\mathtt{BOX2}, ..., \mathtt{BOX125}\}$ and then 
populate the halos within this subvolume with galaxies. 
We implement this procedure to account for sample variance within our forward generative model. 

\subsection{Summary Statistics}\label{sec:statistics}
One of the key ingredients for parameter inference using ABC, is the distance metric between 
the data and the simulations. In essence, it quantifies how close the simulation 
is to reproducing the data. The data and simulation in our scenario (the HOD framework) 
are galaxy populations and their positions. A direct comparison, which would involve comparing 
the actual galaxy positions of the populations, proves to be difficult. 
Instead, a set of statistical summaries are used to encapsulate the information of the data 
and simulations. These quantities should sufficiently describe the information of the data
and simulations while providing the convenience for comparison. 
For the positions of galaxies, sensible summary statistics, 
which we later use in our analysis, include

\begin{itemize}
\item Galaxy number density, $\ngalaxy$: the comoving number density of galaxies computed  
by dividing the comoving volume of the sample from the total number of galaxies. $\ngalaxy$ is 
measured in units of $(\mathrm{Mpc}/h)^{-3}$.  

\item Galaxy two-point correlation function, $\xigg(r)$: a measurement of the excess probability 
of finding a galaxy pair with separation $r$ over an random distribution. To compute $\xigg(rr)$
in our analysis, for computational reasons, we use the Natural estimator (\citealt{peebles80}): 
\beq
\xi(r) = \frac{DD}{RR} - 1, 
\eeq
where $DD$ and $RR$ refer to counts of data-data and random-random pairs.  

\item Galaxy group multiplicity function, $\gmf(N)$: the number density of galaxy groups in bins 
of group richness $N$ where group richness is the number of galaxies within a galaxy group. We 
rely on a Friends-of-Friends (hereafter FoF) group-finder algorithm (\citealt{fof}) to identify 
galaxy groups in our galaxy samples. That is, if the separation of a galaxy pair is smaller than a 
specified linking length, the two galaxies are assigned to the same group. The FoF group-finder 
has been used to identify and analyze the galaxy groups in the SDSS main galaxy sample (\cite{groups}). 
For details regarding the group finding algorithm, we refer readers to \cite{fof}. 

In this study we set the linking length to be $0.25$ times the mean separation of galaxies which 
is given by $\ngalaxy^{-1/3}$. Once the galaxy groups are identified, we bin them into bins of 
group richness. The total number of groups in each bin is divided by the comoving volume to get 
$\gmf(N)$ --- in units of $(\mathrm{Mpc}/h)^{-3}$. 
\end{itemize}

\begin{figure*}
\includegraphics[width=\textwidth]{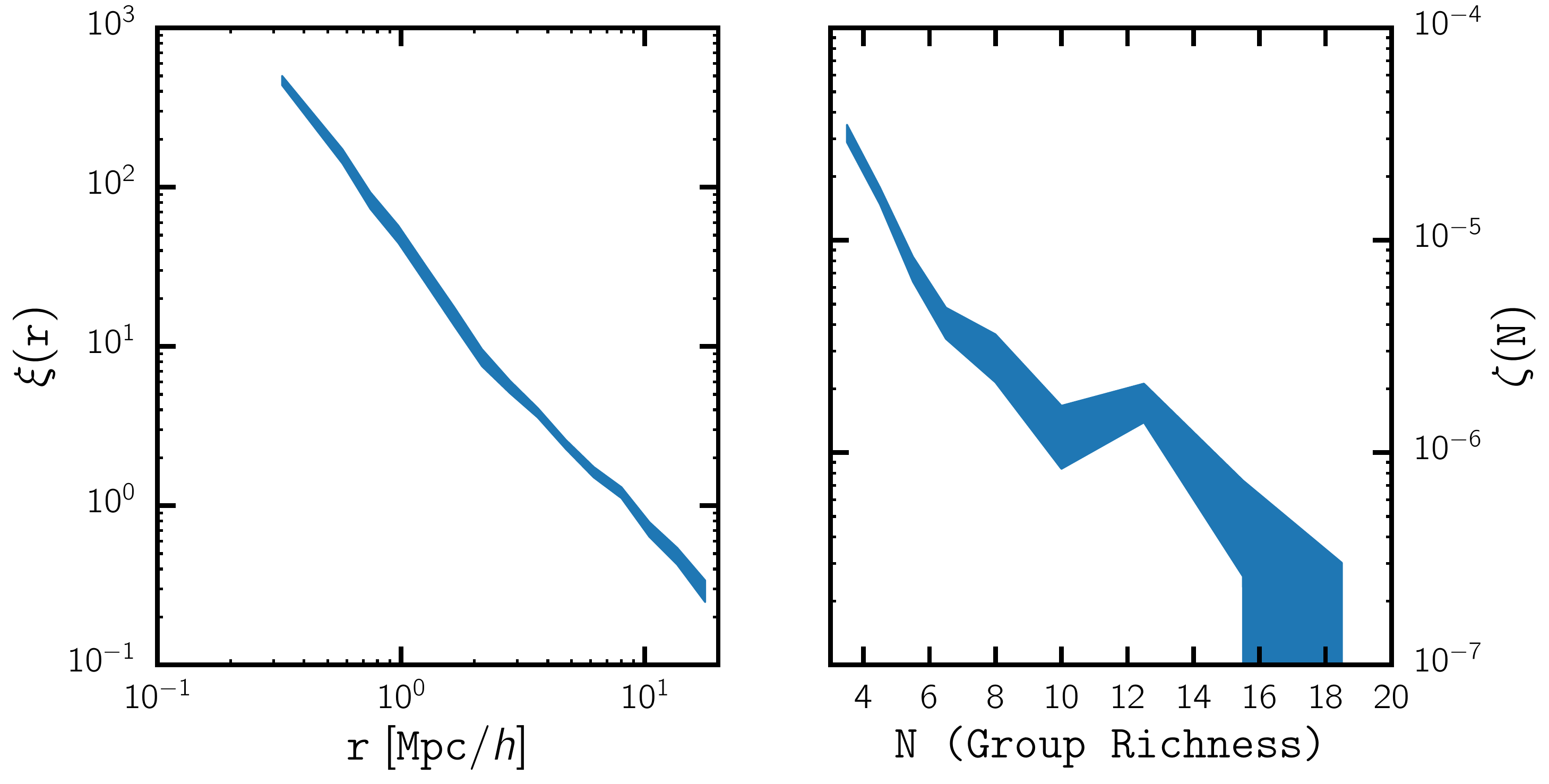}
\caption{\label{fig:mock} The two-point correlation function $\xigg(r)$ (left) and group multiplicity 
function $\gmf(N)$ (right) summary statistics of the mock observations generated from 
the ``true'' HOD parameters described in Section \ref{sec:mock_obv}. The width of the 
shaded region corresponds to the square root of the covariance matrix diagonal 
elements (Eq. \ref{eq:cov}). In our ABC analysis, we treat the $\xigg(r)$ 
and $\gmf(N)$ above as the summary statistics of the observation.}
\end{figure*}

\begin{figure*}
\includegraphics[width=0.8\textwidth]{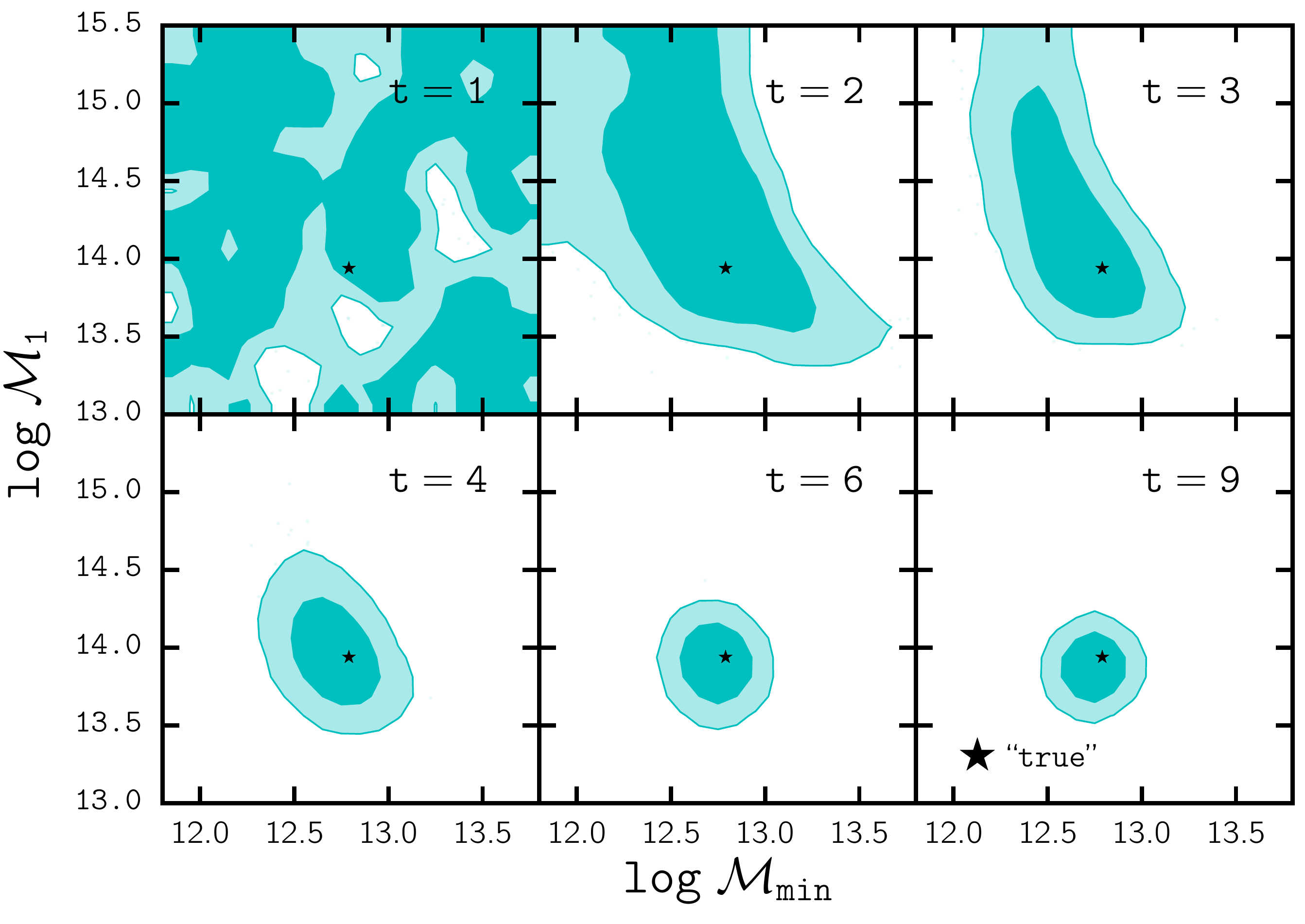}
\caption{\label{fig:pool_demo} We demonstrate the evolution of the ABC particles, $\pars_t$, over iterations $t = 1$ to $9$ in the $\log \mathcal{M}_{min}$ and $\log \mathcal{M}_1$ parameter space. 
$\bar{n}$ and $\gmf(N)$ are used as observables for the above results. For reference, 
in each panel, we include the ``true'' HOD parameters (black star) listed in 
Section \ref{sec:mock_obv}. The initial distance threshold, $\vec\epsilon_1 = 
[\infty, \infty]$ at $t=1$ (top left) so the $\pars_1$ spans the entire range of the prior distribution, 
which is also the range of the panels. We see for $t < 5$, the parameter space occupied 
by the ABC $\pars_t$ shrinks dramatically. Eventually when the algorithm converges, 
$t > 7$, the parameter space occupied by $\pars_t$ no longer shrinks and their 
distributions represent the posterior distribution of the 
parameters. At $t=9$, the final iteration, the ABC algorithm has converged and 
we find that $\pars_\mathrm{true}$ lies safely within the $68\%$ confidence region.} 
\end{figure*}

\section{ABC at work}\label{sec:abcatwork}
With the methodology and the key components of ABC explained above, here we set out to 
demonstrate how ABC can be used to constrain HOD parameters. We start, in Section \ref{sec:mock_obv} 
by creating our ``observation''. We select a set of HOD parameters which we deem as the ``true'' 
parameters and run it through our forward model producing a catalog of galaxy positions 
which we treat as our observation. Then, in Section \ref{sec:abcpmc_spec}, we explain 
the distance metric and other specific choices we make for the ABC-PMC algorithm. 
Ultimately, we demonstrate the use of ABC in LSS, in Section \ref{sec:abc_results},
where we present the parameter constraints we get from our ABC analyses. 
Lastly, in order to both assess the quality of the ABC-PMC parameter inference and also 
discuss the assumptions of the standard Gaussian likelihood approach, we compare the 
ABC-PMC results to parameter constraints using the standard approach in Section 
\ref{sec:abcvsmcmc}.

\subsection{Mock Observations}\label{sec:mock_obv}
In generating our ``observations'', and more generally for our forward model, we adopt 
the HOD model from \cite{zheng07} where the expected number of galaxies populating a 
dark matter halo is governed by Eqs (\ref{eq:ncen}) and (\ref{eq:nsat}). For the parameters 
of the model used to generate the fiducial mock observations, we choose the \cite{zheng07} best-fit HOD parameters for the 
SDSS main galaxy sample with a luminosity threshold $M_{r} = -21$: 
\begin{center}
\renewcommand{\arraystretch}{1.5}
\begin{tabular}{ccccc} \hline \hline 
$\log M_{\rm min}$ & $\sigma_{\log M}$ & $\log M_{0}$ & $\log M_{1}$ & $\alpha$ \\ \hline
$12.79$ & $0.39$ & $11.92$ & $13.94$ & $1.15$ \\ \hline 
\end{tabular} \par
\end{center}
Since these parameters are used to generate the mock observation, they are the parameters
that we ultimately want to recover from our parameter inference. We refer to them as the
true HOD parameters. Plugging them into our forward model (Section \ref{sec:forwardmodel}), 
we generate a catalog of galaxy positions. 

For our summary statistics of the catalogs we use: 
the mean number density $\ngalaxy$, the galaxy two-point correlation function $\xigg(r)$, 
and the group multiplicity function $\gmf(N)$. Our mock observation catalog has 
$\ngalaxy = 9.28875 \times 10^{-4} \;h^{-3} \mathrm{Mpc}^3$ and in Figure \ref{fig:mock} 
we plot $\xigg(r)$ (left panel) and $\gmf(N)$ (right panel). The width of the shaded region
represent the square root of the diagonal elements of the summary statistic covariance matrix, 
which is computed as we describe below. 

We calculate $\xigg$ using the natural estimator (Section \ref{sec:statistics}) with 
fifteen radial bins. The edges of the first radial bin are $0.15$ and $0.5 \; 
h^{-1}\mathrm{Mpc}$. The bin edges for the next 14 bins are logarithmically-spaced between 
$0.5$ and $20\;h^{-1}\mathrm{Mpc}$. We compute the $\gmf(N)$ as described in Section
\ref{sec:statistics} with nine richness bins where the bin edges are logarithmically-spaced 
between $3$ and $20$. To calculate the covariance matrix, we first run the forward model 
using the true HOD parameters for all $125$ halo catalog subvolumes: 
$\{\mathtt{BOX1}, ..., \mathtt{BOX125}\}$. 
We compute the summary statistics of each subvolume galaxy sample $k$: 
\beq
\mathbf{x}^{(k)}=[\ngalaxy, \; \xigg, \; \gmf],
\eeq
Then we compute the covariance matrix as
\begin{eqnarray} 
\mathrm{C}^\mathrm{sample}_{i,j} &=& 
\frac{1}{N_{\mathrm{mocks}}-1}\sum_{k=1}^{N_{\mathrm{mocks}}} 
\Big[\mathbf{x}^{(k)}_{i}-\overline{\mathbf{x}}_{i}\Big]
\Big[\mathbf{x}^{(k)}_{j}-\overline{\mathbf{x}}_{j}\Big], \label{eq:cov} \\
\mathrm{where} \; \overline{\mathbf{x}}_{i} &=& 
\frac{1}{N_{\mathrm{mocks}}}\sum_{k=1}^{N_{\mathrm{mocks}}} \mathbf{x}^{(k)}_{i}.
\end{eqnarray}

Throughout our ABC-PMC analysis, we treat the $\ngalaxy$, $\xigg(r)$, and $\gmf(N)$ we describe in this 
section as if they were the summary statistics of actual observations. 
However, we benefit from the fact that these observables are generated from mock observations 
using the true HOD parameters of our choice: we can use the true HOD parameters to assess the 
quality of the parameter constraints we obtain from ABC-PMC. 

\begin{figure*}
\includegraphics[width=\textwidth]{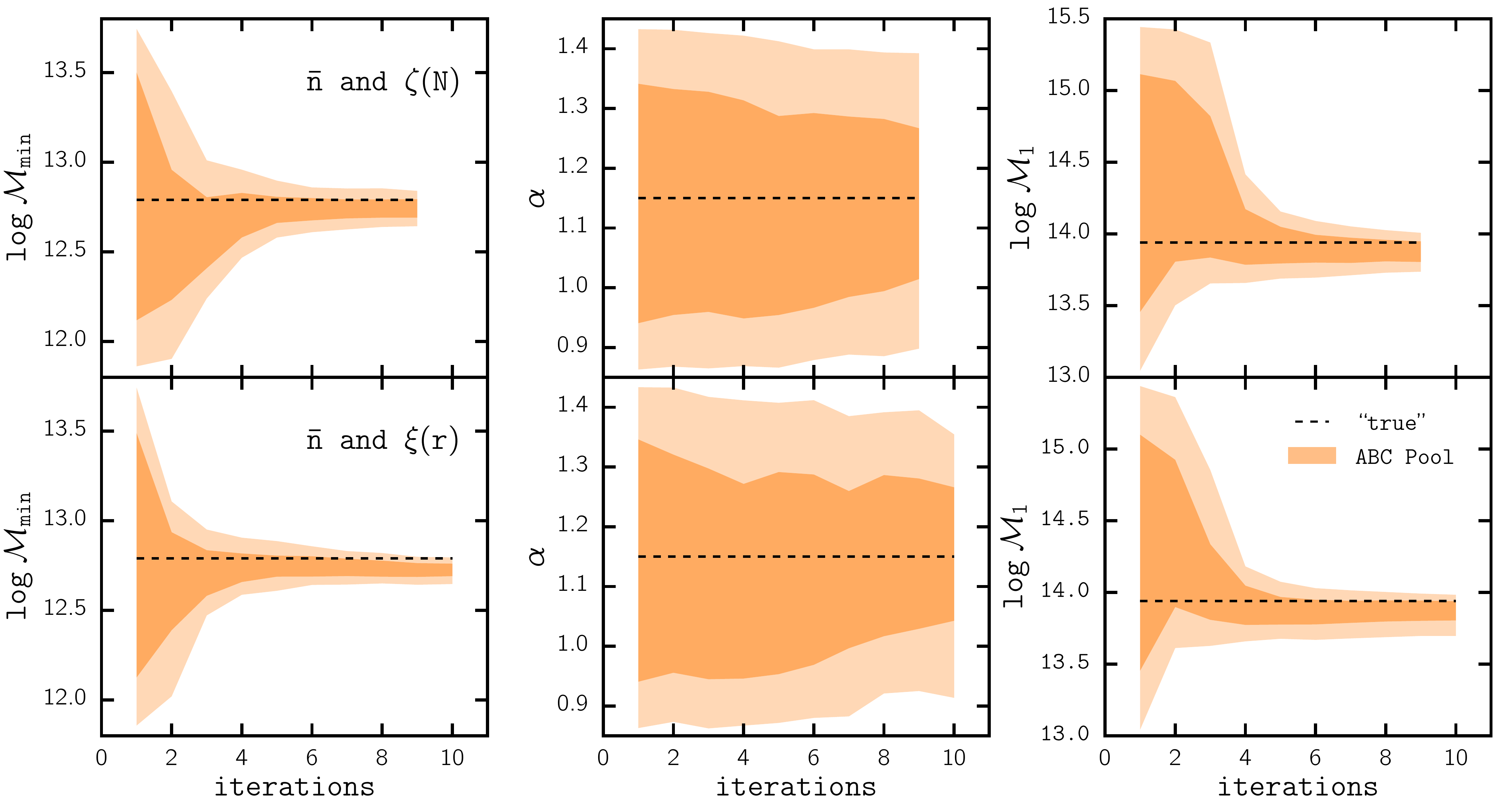}
\caption{\label{fig:abc_converge} We illustrate the convergence of the ABC algorithm through the evolution of the ABC particle distribution as a function of iteration for 
parameters $\log \mathcal{M}_\mathrm{min}$ (left), $\alpha$ (center), and 
$\log \mathcal{M}_1$ (right).
The top panel corresponds our ABC results using the observables $(\bar{n}, \gmf(N))$, 
while the lower panel plots  corresponds to the ABC results using $(\bar{n}, \xigg(r))$. 
The distributions of parameters show no significant change after $t > 7$, which suggests
that the ABC algorithm has converged.}
\end{figure*}

\subsection{ABC-PMC Design} \label{sec:abcpmc_spec}
In Section \ref{sec:abc}, we describe the key components of the ABC algorithm we use in our analysis.
Now, we describe the more specific choices we make within the algorithm: the distance metric, 
the choice of priors, the distance threshold, and the convergence criteria. So far we have 
described three summary statistics: $\ngalaxy$, $\xigg(r)$, and $\gmf(N)$. In order to explore 
the detailed differences in the ABC-PMC parameter constraints based on our choice of summary 
statistics, we run our analysis for two sets of observables: ($\ngalaxy$, $\xigg$) and 
($\ngalaxy$, $\gmf$).

For both analyses, we use a multi-component distance (\citealt{silk12}, Cisewsky et al in preparation). 
Each summary statistic has a distance associated to it: $\rho_{n}$, $\rho_{\xi}$, and $\rho_{\zeta}$. 
We calculate each of these distance components as,   
\begin{eqnarray}
\rho_{n} &=& \frac{\left(\ngalaxy^\mathrm{d} - \ngalaxy^\mathrm{m} \right)^{2}}{\sigma^{2}_{n}}, \label{dn}\\
\rho_{\xi} &=& \sum_{k} \frac{\left[\xigg^\mathrm{d}(r_k) - \xigg^\mathrm{m}(r_k) \right]^{2}}{\sigma_{\xi,k}^{2}}, \label{dxi} \\
\rho_{\zeta} &=& \sum_{k} \frac{\left[\gmf^\mathrm{d}(N_k) - \gmf^\mathrm{m}(N_k) \right]^{2}}{\sigma_{\zeta,k}^{2}}. \label{dgmf}
\end{eqnarray}
The superscripts $\mathrm{d}$ and $\mathrm{m}$ denote the data and model respectively. The data, 
are the observables calculated from the mock observation (Section \ref{sec:mock_obv}). 
$\sigma_{n}^{2}$, $\sigma_{\xi,k}^{2}$, and $\sigma_{\zeta,k}^{2}$ are not the diagonal elements 
of the covariance matrix (\ref{eq:cov}). Instead, they are diagonal elements of the covariance matrix 
$C^\mathrm{ABC}$. 

We construct $C^\mathrm{ABC}$ by populating the entire $\mathtt{MultiDark}$ halo catalogs $125$ times 
repeatedly, calculating $\ngalaxy$, $\xigg$, and $\gmf$ for each realization, and then computing 
the covariance associated with these observables across all realizations. We highlight that $C^\mathrm{ABC}$ 
differs from Eq. \ref{eq:cov}, in that it does not populate the $125$ subvolumes but the entire 
$\mathtt{MultiDark}$ simulation and therefore does not incorporate sample variance. The ABC-PMC analysis 
instead accounts for the sample variance through the forward generative model, which populates the subvolumes
in the same manner as the observations. We use $\sigma_{n}^{2}$, $\sigma_{\xi,k}^{2}$, and $\sigma_{\zeta,k}^{2}$ 
to ensure that the distance is not biased to variations of observables on specific radial or richness bin. 

For our ABC-PMC analysis using the observables $\ngalaxy$ and $\xigg$, our distance metric 
$\vec\rho = [\rho_n, \rho_\xi]$ while the distance metric for the ABC-PMC analysis using 
the observables $\ngalaxy$ and $\gmf$, is $\vec\rho = [\rho_n, \rho_\zeta]$.
To avoid any complications from the choice for our prior, we select uniform priors over all parameters aside from the scatter parameter $\sigma_{\log\;M}$, for which we choose a log-uniform prior. 
We list the range of our prior distributions in Table ~\ref{tab:prior}.

With the distances and priors specified, we now describe the distance thresholds and the 
convergence criteria we impose in our analyses. For the initial iteration, we set distance
thresholds for each distance component to $\infty$. This means, that the initial pool 
$\vec\theta_1$ is simply sampled from the prior distribution we specify above. After the initial 
iteration, the distance threshold is adaptively lowered in subsequent iterations. More 
specifically, we follow the choice of \cite{abcwl} and set the distance threshold 
$\vec\epsilon_t$ to the median of $\vec\rho_{t-1}$, the multi-component distance of the 
previous iteration of particles ($\pars_{t-1}$).

The distance threshold $\vec\epsilon_t$ will progressively decrease. Eventually after a 
sufficient number of iterations, the region of parameter space occupied by $\pars_t$ 
will remain unchanged. As this happens, the acceptance ratio begins to fall significantly. 
When the acceptance ratio drops below $0.001$, our acceptance ratio threshold of choice, 
we deem the ABC-PMC algorithm as converged. In addition to the acceptance ratio threshold 
we impose, we also ensure that distribution of the parameters converges -- another sign 
that the algorithm has converged. Next, we present the results of our ABC-PMC analyses 
using the sets of observables ($\ngalaxy$, $\xigg$) and ($\ngalaxy$, $\gmf$). 


\begin{table}
	\centering
	\caption{{\bf Prior Specifications}: The prior probability distribution 
  and its range for each of the \citet{zheng07} HOD parameters. All mass parameters are in unit of $\mstar$}
	\label{tab:prior}
	\begin{tabular}{lcr} 
		\hline
		HOD Parameter & Prior & Range\\
		\hline
		$\alpha$ & Uniform & [0.8, 1.3]\\
		$\sigma_{\rm \log M}$ & Log-Uniform & [0.1, 0.7]\\
		$\log M_{0}$ & Uniform & [10.0, 13.0]\\
        $\log M_{min}$ & Uniform & [11.02, 13.02]\\
        $\log M_{1}$ & Uniform & [13.0, 14.0]\\
		\hline
	\end{tabular}
\end{table}

\subsection{Results: ABC}\label{sec:abc_results}
We describe the ABC algorithm in Section \ref{sec:abc} and list the particular choices
we make in the implementation in the previous section. Finally, we demonstrate how the 
ABC algorithm produces parameter constraints and present the results of our ABC analysis -- the 
parameter constraints for the \cite{zheng07} HOD model. 

We begin with a qualitative demonstration of the ABC algorithm in Figure \ref{fig:pool_demo},
where we plot the evolution of the ABC $\pars_t$ over the iterations $t = 1$ to $9$, 
in the parameter space of $[\log\mathcal{M}_1, \log\mathcal{M}_{min}]$. The ABC procedure we plot 
in Figure \ref{fig:pool_demo} uses $\bar{n}$ and $\gmf(N)$ for observables, but 
the overall evolution is the same when we use $\bar{n}$ and $\xigg(r)$. The darker and lighter 
contours represent the $68\%$ and $95\%$ confident regions of the posterior distribution over $\pars_t$. For reference, we also plot the ``true'' HOD parameter $\pars_\mathrm{true}$ 
(black star) in each of the panels.
The parameter ranges of the panels are equivalent to the ranges of the prior probabilities we 
specify in Table \ref{tab:prior}. 

For $t=1$, the initial pool (top left), the distance threshold $\vec\epsilon_1 = [\infty , \infty]$, 
so $\pars_1$ uniformly samples the prior probability over the parameters.
At each subsequent iteration, the threshold is lowered (Section \ref{sec:abcatwork}), so for 
$t < 6$ panels, we note that the parameter spaced occupied by $\pars_t$ dramatically shrinks. 
Eventually when the algorithm begins to converge, $t > 7$, the contours enclosing the $68\%$ and 
$95\%$ confidence interval stabilize. At the final iteration $t=9$ (bottom right), 
the algorithm has converged and we find that 
$\pars_\mathrm{true}$ lies within the $68\%$ confidence interval of the $\pars_{t=9}$ particle 
distribution. This $\pars_t$ distribution at the final iteration represents the 
posterior distribution of the parameters. 

To better illustrate the criteria for convergence, in Figure \ref{fig:abc_converge}, 
we plot the evolution of the $\pars_t$ distribution as a function of iteration 
for parameters $\log\mathcal{M}_\mathrm{min}$ (left), $\alpha$ (center), and 
$\log\mathcal{M}_1$ (right). The darker and lighter shaded regions correspond to the 
$68\%$ and $95\%$ confidence levels of the $\pars_t$ distributions. The top panels 
correspond to our ABC results using $(\bar{n}, \gmf)$ as observables and the bottom 
panels correspond to our results using $(\bar{n}, \xigg)$. 
For each of the parameters in both top and bottom panels, we find that the distribution 
does not evolve significantly for $t > 7$. At this point additional iterations in
our ABC algorithm will neither impact the distance threshold $\vec\epsilon_t$ nor 
the posterior distribution of $\pars_t$. We also emphasize that the convergence of the 
parameter distributions coincides with when the acceptance ratio, discussed in Section 
\ref{sec:abcpmc_spec}, crosses the predetermined shut-off value of $0.001$. Based on these criteria, 
our ABC results for both $(\bar{n}, \gmf)$ and $(\bar{n}, \xigg)$ observables have converged. 

\begin{figure*}
\includegraphics[width=0.85\textwidth]{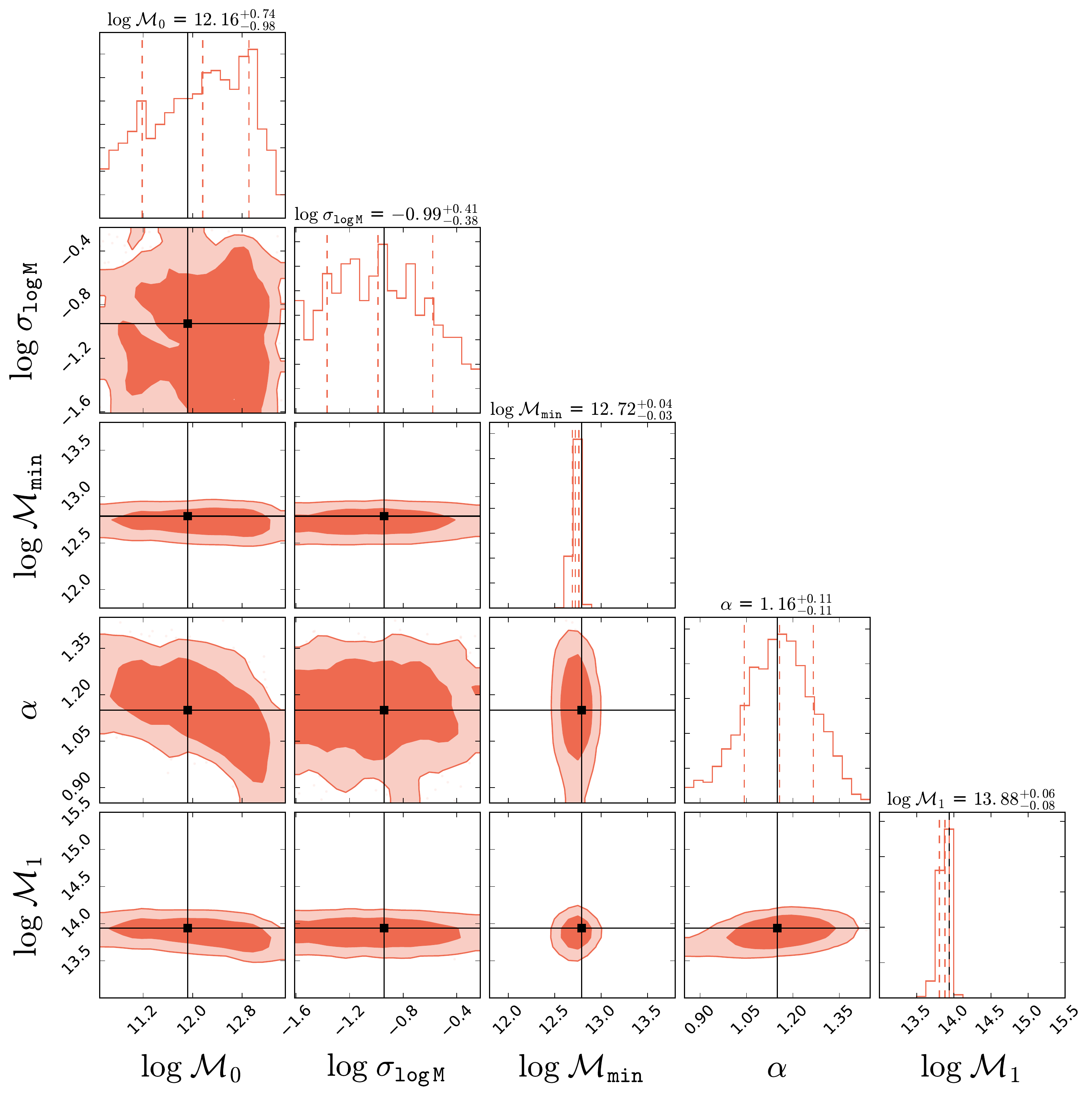}
\caption{\label{fig:abc_corner_nbarxi} We present the constraints on the \citet{zheng07} HOD model parameters obtained from our ABC-PMC analysis using $\bar{n}$ and $\xigg(r)$ as observables.  
The diagonal panels plot the posterior distribution of each HOD parameter with 
vertical dashed lines marking the $50\%$ quantile and $68\%$ confidence intervals of the 
distribution. The off-diagonal panels plot the degeneracies between parameter pairs. 
The range of each panel corresponds to the range of our prior choice. The ``true''
HOD parameters, listed in Section \ref{sec:mock_obv}, are also plotted in each of 
the panels (black). For $\log\mathcal{M}_0$, $\alpha$, and $\sigma_{\log M}$, 
the ``true'' parameter values lie near the center of the $68\%$ confidence interval 
of the posterior distribution. For $\log\mathcal{M}_1$ and $\log\mathcal{M}_\mathrm{min}$, 
which have tight constraints, the ``true'' values lie within the $68\%$ confidence 
interval. Ultimately, the ABC parameter constraints we obtain in our analysis are consistent with the ``true'' HOD parameters.}
\end{figure*}

\begin{figure*}
\includegraphics[width=0.85\textwidth]{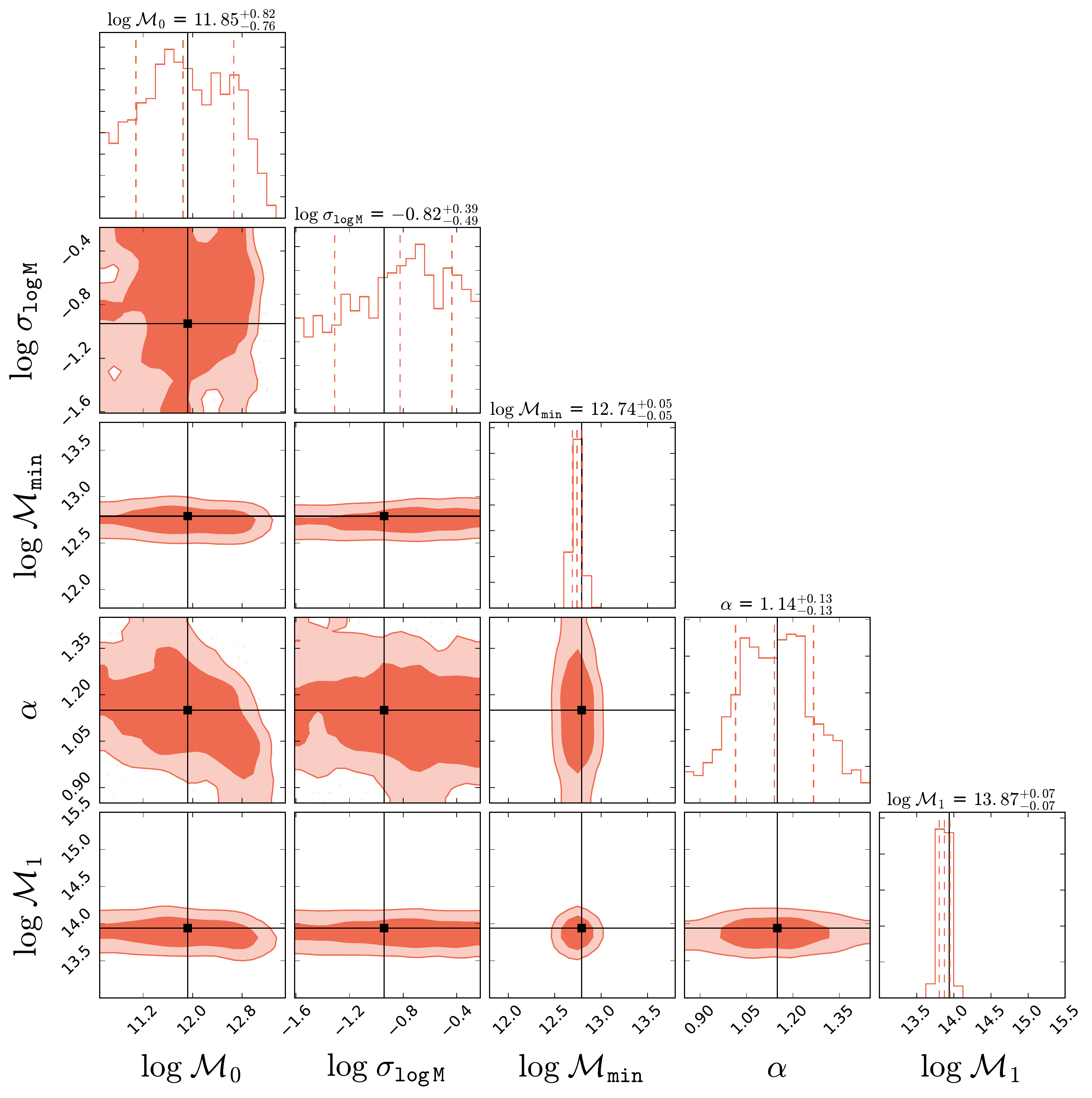}
\caption{\label{fig:abc_corner_nbargmf}Same as Figure \ref{fig:abc_corner_nbarxi} but for our ABC analysis using $\bar{n}$ and $\gmf(N)$ as observables. The ABC parameter constraints we obtain are
consistent with the ``true'' HOD parameters.}
\end{figure*}

We present the parameter constraints from the converged ABC analysis in Figure \ref{fig:abc_corner_nbarxi} and 
Figure \ref{fig:abc_corner_nbargmf}. Figure \ref{fig:abc_corner_nbarxi} shows 
the parameter constraints using $\bar{n}$ and $\xigg(r)$ while Figure 
\ref{fig:abc_corner_nbargmf} plots the constraints using $\bar{n}$ and $\gmf(N)$.
For both figures, the diagonal panels plot the posterior distribution of the 
HOD parameters with vertical dashed lines marking the $50\%$ (median) and $68\%$ 
confidence intervals. The off-diagonal panels plot the degeneracy between  
parameter pairs. To determine the accuracy of our ABC parameter constraints, we plot
the ``true'' HOD parameters (black) in each of the panels.
For both sets of observables, our ABC constraints are consistent with the 
``true'' HOD parameters. For $\log\mathcal{M}_0$, $\log\sigma_{\log M}$,
and $\alpha$, the true parameter values lie near the center of the $68\%$ confidence 
interval. For the other parameter, which have much tighter constraints, the true 
parameters lie within the $68\%$ confidence interval.  

\begin{figure*}
\includegraphics[width=\textwidth]{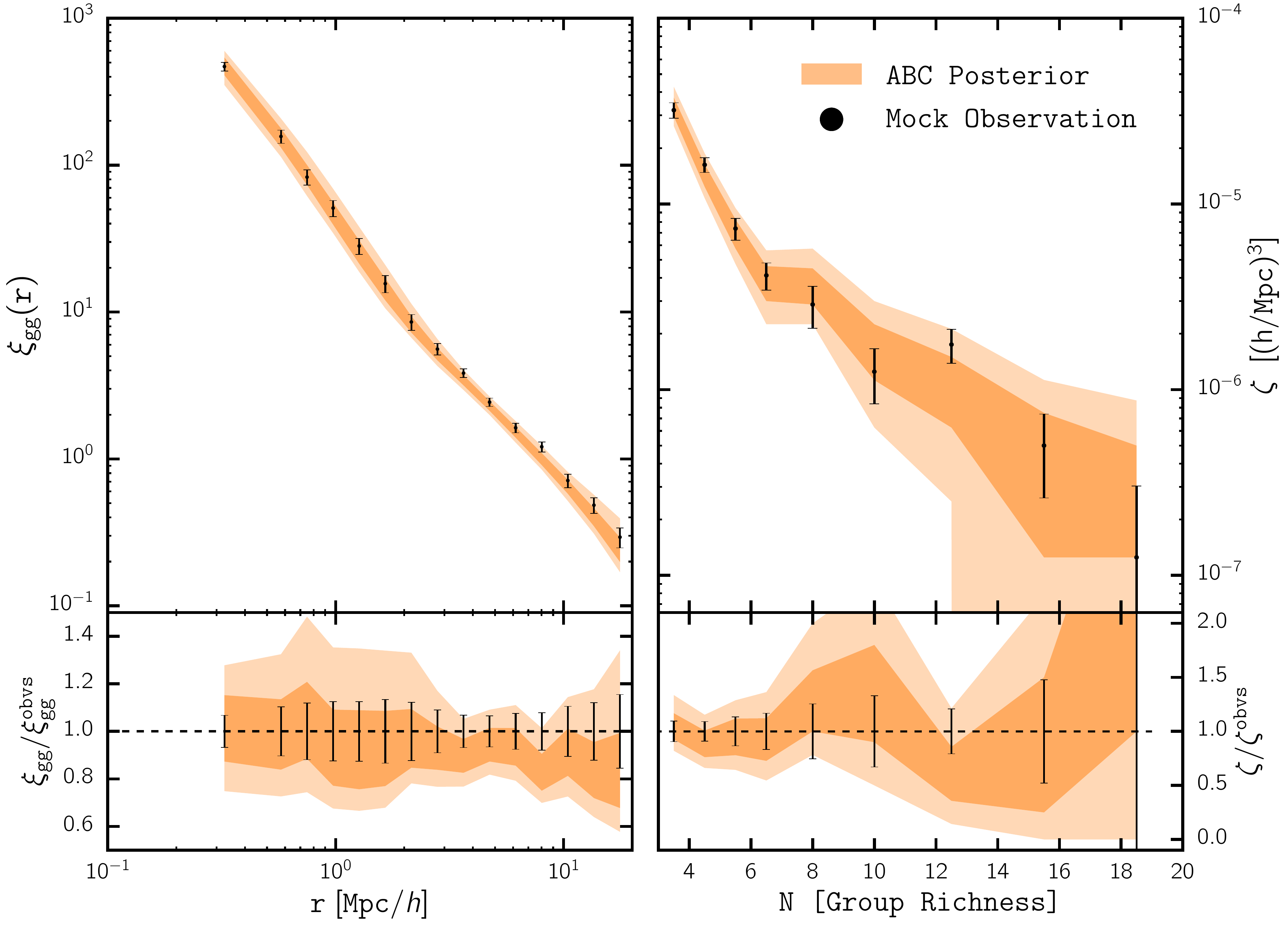}
\caption{\label{fig:abc_moneyplot} We compare the ABC-PMC posterior prediction for the observables $\xigg(r)$ (left) and $\gmf(N)$ (right) (orange; Section \ref{sec:abc_results}) 
to $\xigg(r)$ and $\gmf(N)$ of the mock observation (black) in the top panels.
In the lower panels, we plot the ratio between the ABC-PMC posterior predictions for $\xigg$ and $\gmf$ 
to the mock observation $\xigg^\mathrm{obvs}$ and $\gmf^\mathrm{obvs}$. The darker 
and lighter shaded regions represent the $68\%$ and $95\%$ confidence regions of the posterior predictions, 
respectively. The error-bars represent the square root of the diagonal elements of the error 
covariance matrix (equation \ref{eq:cov}) of the mock observations. Overall, the observables 
drawn from the ABC-PMC posteriors are in good agreement with $\xigg$ and $\gmf$ of the 
mock observations. The lower panels demonstrate that for both observables, the error-bars 
of the mock observations lie within the $68\%$ confidence interval of the ABC-PMC posterior 
predictions.} 
\end{figure*}

To further test the ABC results, in Figure \ref{fig:abc_moneyplot}, 
we compare $\xigg(r)$ (left) and $\gmf(N)$ (right) of the mock observations from Section \ref{sec:mock_obv}
to the predictions of the ABC posterior distribution (shaded). The error bars of the mock observations 
represent the square root of the diagonal elements of the covariance matrix (Eq. \ref{eq:cov}) while the 
darker and lighter shaded regions represent the $68\%$ and $95\%$ confidence regions of the ABC posterior 
predictions. In the lower panels, we plot the ratio of the ABC posterior prediction $\xigg(r)$ and $\gmf(N)$ over
the mock observation $\xigg^\mathrm{obvs}(r)$ and $\gmf^\mathrm{obvs}(N)$. Overall, the 
ratio of the $68\%$ confidence region of ABC posterior predictions is consistent with unity 
throughout the $r$ and $N$ range. We observe slight deviations in the $\xigg$ ratio for 
$r > 5\;\mathrm{Mpc}/h$; however, any deviation is within the uncertainties of the mock observations. 
Therefore, the observables drawn from the ABC posterior distributions are in good agreement with the
observables of the mock observation. 

The ABC results we obtain using the algorithm of Section \ref{sec:abc} with the choices of
Section \ref{sec:abcpmc_spec} produce parameter constraints that are consistent with 
the ``true'' HOD parameters (Figures \ref{fig:abc_corner_nbarxi} and \ref{fig:abc_corner_nbargmf}).
They also produce observables $\xigg(r)$ and $\gmf(N)$ that are consistent with $\xigg^\mathrm{obvs}$ 
and $\gmf^\mathrm{obvs}$. Thus, through ABC we are able to produce consistent 
parameter constraints. {\em More importantly, we demonstrate that ABC is feasible 
for parameter inference in large scale structure.} 

\subsection{Comparison to the Gaussian Pseudo-Likelihood MCMC Analysis} \label{sec:abcvsmcmc}
In order to assess the quality of the parameter inference described in the previous section, 
we compare the ABC-PMC results with the HOD parameter constraints from
assuming a Gaussian likelihood 
function. The model used for the Gaussian likelihood analysis is different than the forward 
generative model adopted for the ABC-PMC algorithm, to be consistent with the standard approach.

In the ABC analysis, the model accounts for sample variance by randomly sampling a subvolume to be 
populated with galaxies. Instead, in the Gaussian pseudo-likelihood analysis, the covariance matrix is assumed to capture the
uncertainties from sample variance. Hence, in the model for the Gaussian pseudo-likelihood analysis, 
we populate halos of the {\em entire} $\mathtt{MultiDark}$ simulation rather than a subvolume.
We describe the Gaussian pseudo-likelihood analysis below.


To write down the Gaussian pseudo-likelihood, we first introduce the vector $\mathbf{x}$: 
a combination of the summary statistics (observables) for a galaxy catalog. 
When we use $\ngalaxy$ and $\xigg(r)$ as observables in the analysis: $\mathbf{x} = [\ngalaxy , \xigg]$;
when we use $\ngalaxy$ and $\gmf(N)$ as observables in the analysis: $\mathbf{x} = [\ngalaxy , \gmf]$.
Based on this notation, we can write pseudo-likelihood function as 
\begin{eqnarray}
-2 \ln \mathcal{L}(\theta | d) &=& \Delta \mathbf{x}^{T}\widehat{C^{-1}}\Delta \mathbf{x} + \ln\Big[(2\pi)^{d}\mathrm{det}(C)\Big], \label{loglike}
\end{eqnarray}
where 
\begin{eqnarray}
\Delta \mathbf{x} &=& [\mathbf{x}_{obs} -\mathbf{x}_{mod}], 
\end{eqnarray}
the difference between $\mathbf{x}_{obs}$, measured from the mock observation, 
and $\mathbf{x}_{mod}(\mathbb{\theta})$ measured from the mock catalog generated 
from the model with parameters $\theta$ .
$d$ here is the dimension of $\mathbf{x}$ (for $\mathbf{x} = [\ngalaxy, \xigg]$, $d = 13$; 
for $\mathbf{x} = [\ngalaxy, \gmf]$, $d = 10$).  
$\widehat{C^{-1}}$ is the inverse covariance matrix, which we estimate following \cite{hartlap2007}:
\begin{eqnarray}
\widehat{C^{-1}} = \frac{N_{\rm mocks}-d - 1}{N_{\rm mocks} -1} \; \widehat{C}^{-1}.
\end{eqnarray}
$\widehat{C}$ is the estimated covariance matrix, calculated using the corresponding 
$\mathbf{x}$ block of the covariance matrix from Eq.~\ref{eq:cov}, and $N_{\rm mock}$ 
is the number of mocks used for the estimation ($N_{\rm mock} = 124$; see Section \ref{sec:mock_obv}).
We note that in $\widehat{C}$ the dependence on the HOD parameters is neglected, 
so the second term in the expression of Eq.~\ref{loglike} can be neglected. 
Finally, using this pseudo-likelihood, we sample from the posterior distribution 
given the prior distribution using the MCMC sampler $\mathtt{emcee}$ (\citealt{emcee}). 

\begin{figure*}
\includegraphics[width=1.\textwidth]{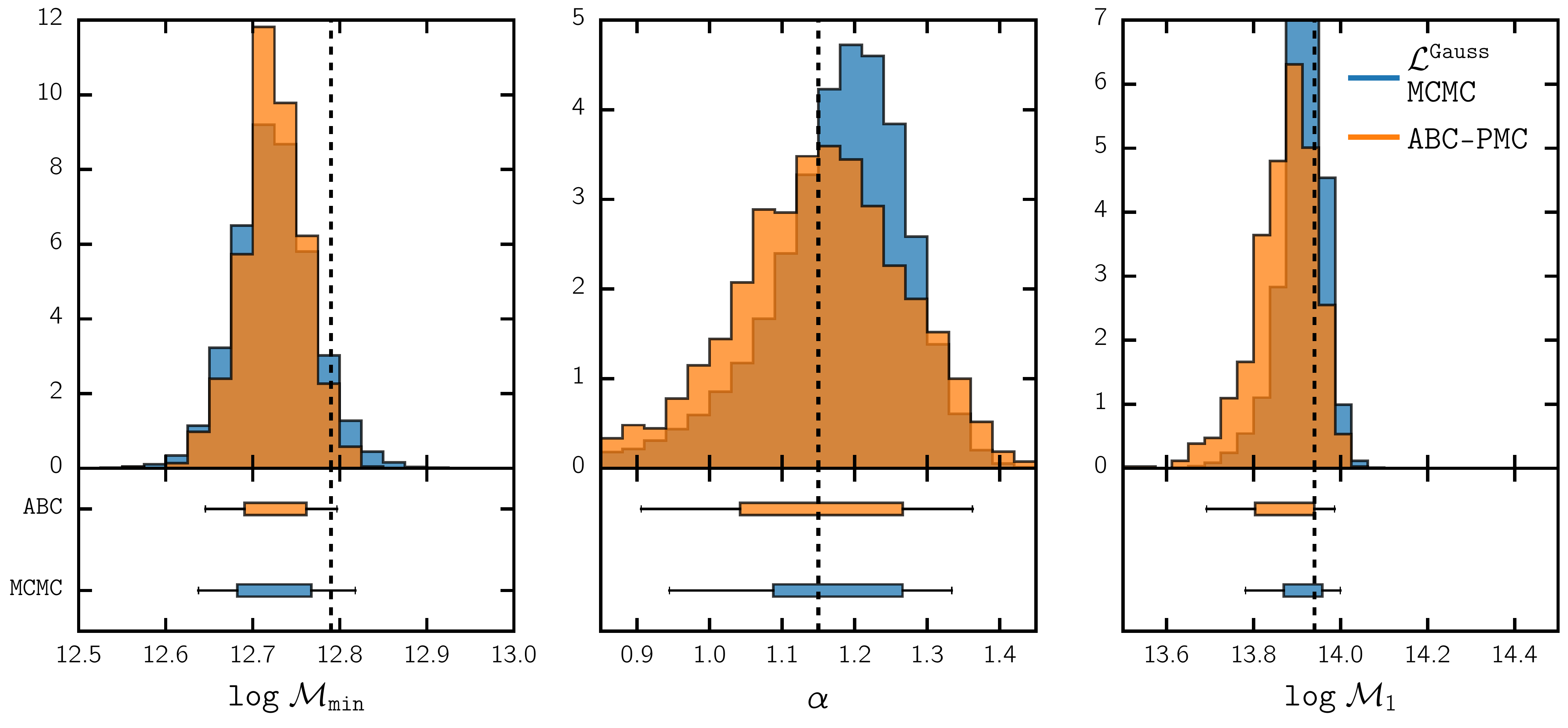}
\caption{\label{fig:hist_nbarxi} 
We compare the $\log \mathcal{M}_{\rm min}$, $\alpha$, and $\log \mathcal{M}_{1}$ parameter 
constraints from ABC-PMC (orange) to constraints from the Gaussian pseudo-ikelihood MCMC (blue) 
using $\ngalaxy$ and $\xigg(r)$ as observables. The \emph{top} panels compares the two methods' 
marginalized posterior PDFs over the parameters. In the \emph{bottom} panels, we include box 
plots marking the confidence intervals of the posterior distributions. The boxes represent the 
$68\%$ confidence interval while the ``whiskers'' represent the $95\%$ confidence interval. We mark 
the ``true'' HOD parameters with vertical black dashed line. The marginalized posterior PDFs 
obtained from the two methods are consistent with each other. The ABC-PMC and Gaussian pseudo-likelihood
constraints are generally consistent for $\log \mathcal{M}_{\rm min}$ and $\log \mathcal{M}_{1}$. 
The ABC-PMC constraint for $\alpha$ is slightly less biased and has slightly larger uncertainty
then the constraint from Gaussian pseudo-likelihood analysis.} 
\end{figure*}

\begin{figure*}
\includegraphics[width=1.\textwidth]{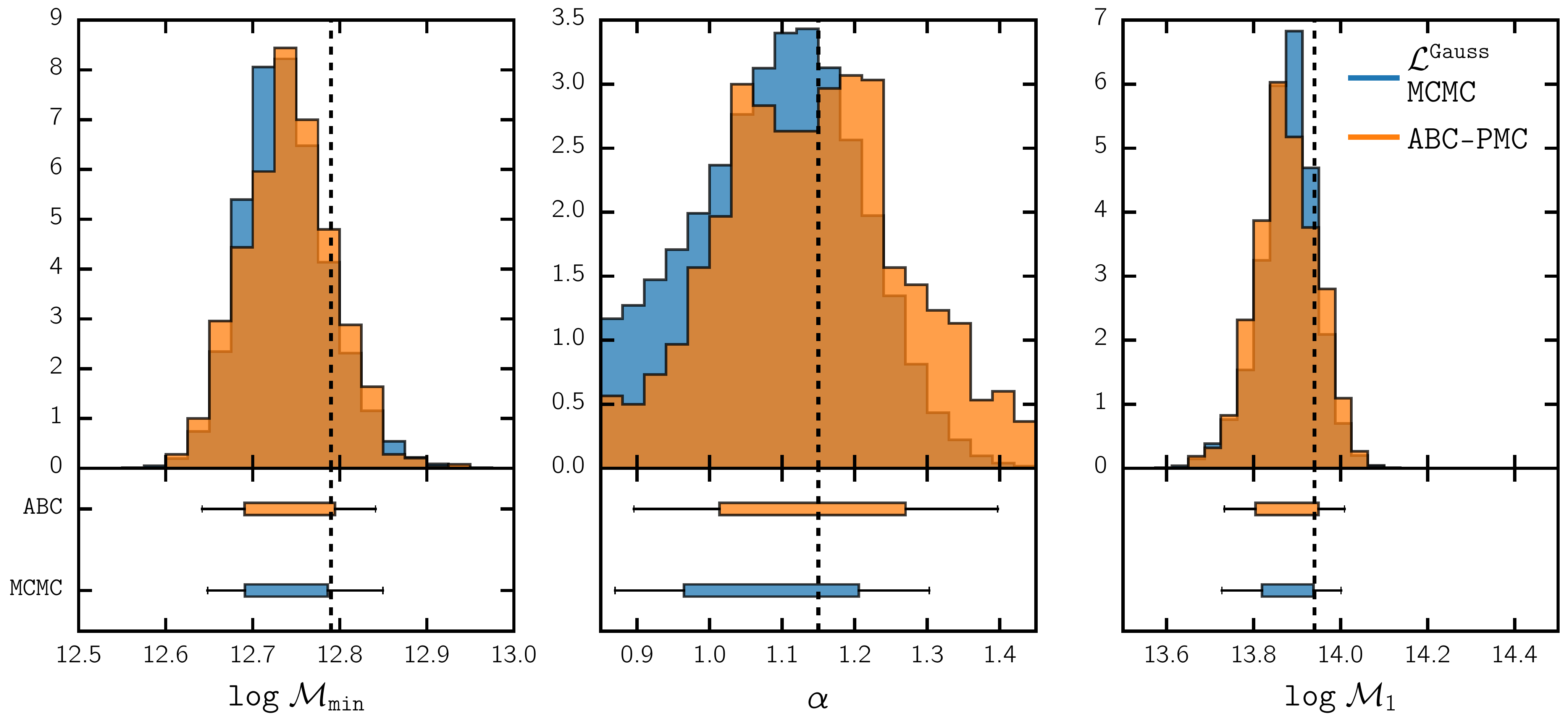}
\caption{\label{fig:hist_nbargmf} 
Same as Figure \ref{fig:hist_nbarxi}, but both the ABC-PMC
analysis and the Gaussian pseudo-likelihood MCMC analysis use $\ngalaxy$ and $\gmf(N)$ as 
observables. Both methods derive constraints consistent with the ``true'' HOD parameters 
and infer the region of allowed values to similar precision. We note that the MCMC constraint 
on $\alpha$ is slightly more biased compared to ABC-PMC estimate. This discrepancy may stem 
from the fact that the use of Gaussian pseudo-likelihood and its associated assumptions is more 
spurious when modeling the group multiplicity function.}
\end{figure*}

In Figures~\ref{fig:hist_nbarxi} and \ref{fig:hist_nbargmf}, we compare the results from
ABC-PMC and Gaussian pseudo-likelihood MCMC analyses using $[\ngalaxy, \xigg]$ and 
$[\ngalaxy, \gmf]$ as observables, respectively. The top panels in each figure 
compares the marginalized posterior PDFs for three parameters of the HOD model: 
$\{\log \mathcal{M}_{\rm min}, \alpha, \log\mathcal{M}_1\}$. The lower panels in each 
figure compares the $68\%$ and $95\%$ confidence intervals of the constraints 
derived from the two inference methods as a box plot. The ``true'' HOD parameters
are marked by vertical dashed lines in each panel.

In both Figures~\ref{fig:hist_nbarxi} and \ref{fig:hist_nbargmf}, the marginalized 
posteriors for each of the parameters from both inference methods are comparable 
and consistent with the ``true'' HOD parameters. However, we note that there are 
minor discrepancies between the maringalized posterior distributions. In particular, 
the distribution for $\alpha$ derived from ABC-PMC is less biased than the $\alpha$ 
constraints from the Gaussian pseudo-likelihood approach. 

\begin{figure*}
\includegraphics[width=1.\textwidth]{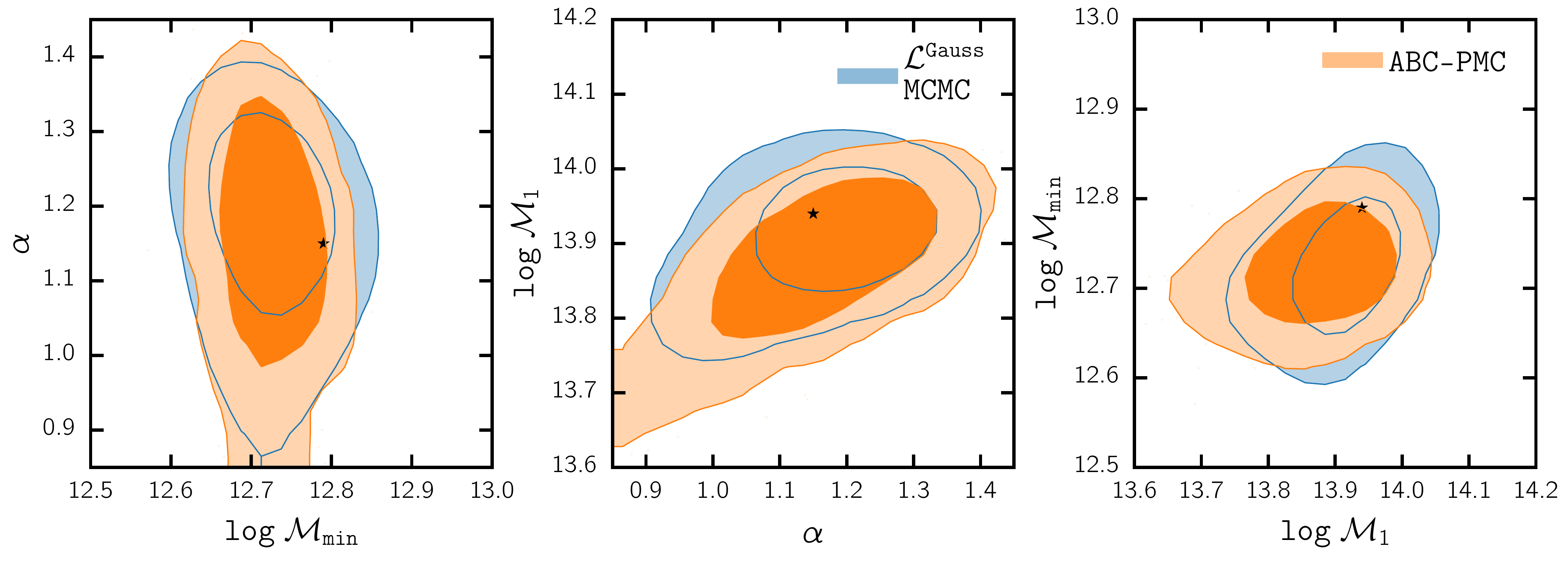}
\caption{\label{fig:cont_nbarxi}
We compare the ABC-PMC (orange) and the Gaussian pseudo-likelihood MCMC (blue)
predictions of the 68\% and 95\% posterior confidence regions over the HOD 
parameters ($\log \mathcal{M}_{\rm min}$, $\alpha$, and $\log \mathcal{M}_{1}$) 
using $\ngalaxy$ and $\xigg(r)$ as observables. In each panel, the black star
represents the ``true'' HOD parameters used to generate the mock observations. 
Both inference methods derive confidence regions consistent with the ``true'' 
HOD parameters.}
\end{figure*}
\begin{figure*}
\includegraphics[width=1.\textwidth]{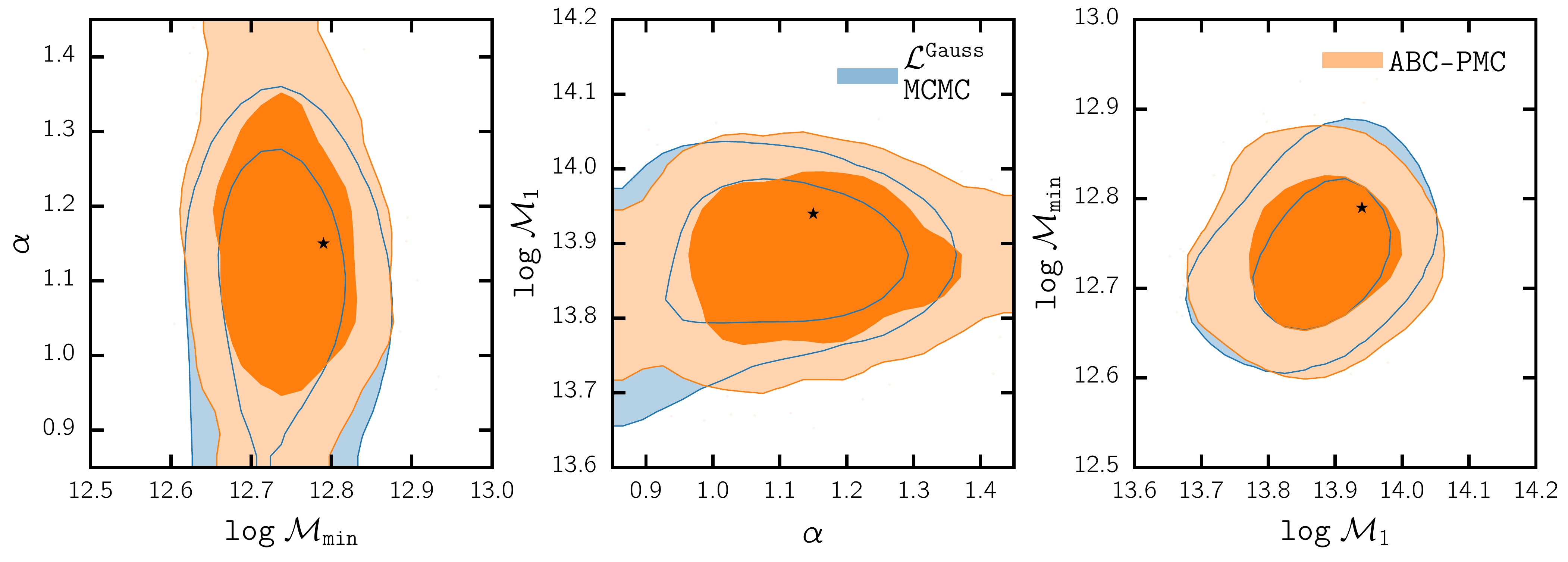}
\caption{\label{fig:cont_nbargmf} 
Same as Figure \ref{fig:cont_nbarxi}, but using $\ngalaxy$ and $\gmf(N)$ as 
observables. Again, the confidence regions derived from both methods are 
consistent with the ``true'' HOD parameters used to generate the mock 
observations. The confidence region of $\alpha$ from the Gaussian 
pseudo-likelood method is biased compared to the ABC-PMC contours. 
This may be due to the fact that the true likelihood function that 
describes $\gmf(N)$ deviates significantly from the assumed Gaussian 
functional form.}
\end{figure*}

In Figures \ref{fig:cont_nbarxi} and \ref{fig:cont_nbargmf}, we plot the contours 
enclosing the $68\%$ and $95\%$ confidence regions of the posterior probabilities of 
the two methods using $[\ngalaxy, \xigg]$ and $[\ngalaxy, \gmf]$ as observables 
respectively. In both figures, we mark the ``true'' HOD parameters (black star). The 
overall shape of the contours are in agreement with each other. However, we note
that the contours for the ABC-PMC method are more extended along $\alpha$. 


Overall, the HOD parameter constraints from ABC-PMC are consistent with those from 
the Gaussian pseudo-likelihood MCMC method; however, using ABC-PMC has a number of 
advantages. For instance, ABC-PMC utilizes a forward generative model. Our forward 
generative model accounts for sample variance. On the other hand, the Gaussian 
pseudo-likelihood approach, as mentioned earlier this section, does not account 
for sample variance in the model and relies on the covariance matrix estimate to 
capture the sample variance of the data.

Accurate estimation of the covariance matrix 
in LSS, however, faces a number of challenges. It is both labor and computationally 
expensive and dependent on the accuracy of simulated mock catalogs, known to be 
unreliable on small scales~(see \citealt{cosmiccode,nifty} and references therein). 
In fact, as \cite{Sellentin:2016a} points out, using estimates of the covariance 
matrix in the Gaussian psuedo-likelihood approach become further problematic. Even 
when inferring parameters from a Gaussian-distributed data set, using covariance 
matrix estimates rather than the {\em true} covariance matrix leads to a likelihood 
function that is {\em no longer} Gaussian. ABC-PMC does not depend on a covariance 
matrix estimate; hence, it does not face these problems.

In addition to not requiring accurate covariance matrix estimates, forward models 
of the ABC-PMC method, in principle, also have the advantage that they can account 
for sources of systematic uncertainties that affect observations. All observations 
suffer from significant systematic effects which are often difficult to correct. 
For instance, in SDSS-III BOSS~\citep{Dawson:2013a}, fiber collisions and redshift
failures siginifcantly bias measurements and analysis of observables such as 
$\xigg$ or the galaxy powerspectrum~\citep{Ross:2012aa, Guo:2012a, Hahn:2017a}. 
In parameter inference, these systematics can affect the likelihood, and thus any 
analysis that requires writing down the likelihood, in unknown ways. With a forward generative model of the ABC-PMC 
method, the systematics can be simulated and marginalized out to achieve unbiased 
constraints. 


Furthermore, {\em ABC-PMC -- unlike the Gaussian pseudo-likelihood approach -- 
is agnostic about the functional form of the underlying distribution of the 
summary statistics} (\emph{e.g.} $\xigg$ and $\gmf$). As we explain throughout 
the paper, the likelihood function in LSS {\em cannot} be Gaussian. For $\xigg$, 
the correlation function must satisfy non-trivial positive-definiteness requirements 
and hence the Gaussian pseudo-likelihood function assumption is not correct 
in detail. In the case of $\gmf(N)$, assuming a Gaussian functional form for 
the likelihood, which in reality is more likely Poisson, misrepresents the true 
likelihood function. In fact, this incorrect 
likelihood, may explain why the constraints on $\alpha$ are less biased for 
the ABC-PMC analysis than the Gaussian-likelihood analysis in \ref{fig:cont_nbargmf}.


Although in our comparison using simple mock observations, we find generally 
consistent parameter constraints from both the ABC-PMC analysis and the standard 
Gaussian pseudo-likelihood analysis, more realistic scenarios present many factors
that can generate inconsistencies. Consider a typical galaxy catalog from 
LSS observations. These catalogs consist of objects with different data 
qualities, signal-to-noise ratios, and systematic effects. For example, catalogs 
are often incomplete beyond some luminosity/redshift or have some threshold 
signal-to-noise ratio cut imposed on them.


These selection effects, coupled with the systematic effects earlier this section, 
make correctly predicting the likelihood intractable. In the standard Gaussian 
pseudo-likelihood analysis, and other analysis that require writing down a 
likelihood function, these effects can significantly 
bias the inferred parameter constraints. In these situations, employing
ABC equipped with a generative forward model that incorproates selection and 
systematic effects may produce less biased parameter constraints. 


Despite the advantages of ABC, one obstactle for adopting it to parameter 
inference has been the computational costs of generative forward models, a 
key element of ABC. By combining ABC with the PMC sampling method, however, 
ABC-PMC efficiently converges to give reliable posterior parameter constraints. 
In fact, in our analysis, the total computational resources required for the 
ABC-PMC analysis were {\em comparable} to the computational resources used 
for the Gaussian pseudo-likelihood analysis with MCMC sampling.

Applying ABC-PMC beyond the analysis in this work, to broader LSS analyses 
imposes some caveats. In this work, we focus on the galaxy-halo connection, so
our generative forward model populates halos with galaxies. LSS analyses 
for inferring cosmological parameters would require generating halos by 
running cosmological simulations. The forward models also need to accurately 
model the observation systematic effects of the latest observations. Hence, 
accurate generative forward models in LSS analyses demand improvements in simulations 
and significant computational resources in order to infer unbiased parameter 
constraints. Recent cosmology simulations show promising improvements 
in both accuracy and speed~\citep[\emph{e.g.}][]{fastpm}. Such developements
will be crucial for applying ABC-PMC to broader LSS analyses and exploiting 
the significant advantages that ABC-PMC offers.


\section{Summary and Conclusion}\label{sec:discussion}

Approximate Bayesian Computation, ABC, is a generative, simulation-based
inference that can deliver correct parameter estimation with
appropriate choices for its design.
It has the advantage over the standard approach in that it 
does not require explicit knowledge of the likelihood function. It only relies on the ability to simulate
the observed data, accounting for the uncertainties associated with observation and on specifying a metric 
for the distance between the observed data and simulation. When the specification of the likelihood function 
proves to be challenging or when the true underlying distribution of the observable is unknown, 
ABC provides a promising alternative for inference.

The standard approach to large scale structure studies relies on the assumption that the likelihood function 
for the observables -- often two-point correlation function -- given the model has a Gaussian functional form. 
In other words, it assumes that the statistical summaries are Gaussian distributed. In principle to rigorously 
test such an assumption, a large number of realistic simulations would need to be generated in order to examine 
the actual distribution of the observables. This process, however, is prohibitively---both labor and computationally 
---expensive. Therefore, our assumption of a Gaussian likelihood function remains largely unconfirmed and so 
unknown. Fortunately, the framework of ABC permits us to bypass any 
assumptions regarding the distribution of observables. Through ABC, we can provide constraints for our models without making the unexamined assumption of Gaussianity. 

With the ultimate goal of demonstrating that ABC is feasible for LSS studies, we use it to 
constrain parameters of the halo occupation distribution, which dictates the galaxy-halo connection. 
We begin by constructing a mock observation of galaxy distribution with a chosen set of ``true'' HOD model 
parameters. Then we attempt to constrain these parameters using ABC. More specifically, in this paper: 

\begin{itemize}
\item We provide an explanation of the ABC algorithm and present how Population Monte Carlo can be utilized 
to efficiently reach convergence and estimate the posterior distributions of model parameters. 
We use this ABC-PMC algorithm with a generative forward model built with $\mathtt{Halotools}$, a software 
package for creating catalogs of galaxy positions based on models of the galaxy-halo connection such as the HOD. 

\item We choose $\ngalaxy$, $\xigg$ and $\gmf$ as observables and summary statistics of the galaxy position catalogs. 
And for our ABC-PMC algorithm, we specify a multi-component distance metric, uniform priors, 
a median threshold implementation, and an acceptance rate-based convergence criterion.

\item From our specific ABC-PMC method, we obtain parameter constraints that are consistent with the ``true'' HOD parameters of our mock observations. Hence we demonstrate that ABC-PMC can be used for parameter inference in LSS studies. 

\item We compare our ABC-PMC parameter constraints to constraints using the standard Gaussian-likelihood
MCMC analysis. The constraints we get from both methods are comparable in accuracy and precision. However, 
for our analysis using $\ngalaxy$ and $\gmf$ in particular, we obtain less biased posterior distributions when
comparing to the ``true'' HOD parameters. 
\end{itemize}

Based on our results, we conclude that ABC-PMC is able to consistently infer parameters in the context
of LSS. We also find that the computation required for our ABC-PMC and standard Gaussian-likelihood 
analyses are comparable. Therefore, with the statistical advantages that ABC offers, we present ABC-PMC 
as an improved alternative for parameter inference.

\section*{Acknowledgements}

We thank Jessie Cisewsky for reading and making valuable comments on the draft. We would also like to thank Michael R. Blanton, Jeremy R. Tinker, Uros Seljak, Layne Price, 
Boris Leidstadt, Alex Malz, Patrick McDonald, and Dan Foreman-Mackey for productive 
and insightful discussions. MV was supported by NSF grant AST-1517237. DWH was supported 
by NSF (grants IIS-1124794 and AST-1517237), NASA (grant NNX12AI50G), and the Moore-Sloan 
Data Science Environment at NYU. KW was supported by NSF grant AST-1211889. Computations 
were performed using computational resources at NYU-HPC. We thank Shenglong Wang, the 
administrator of NYU-HPC computational facility, for his consistent and continuous support 
throughout the development of this project. We would like to thank the organizers of 
the AstroHackWeek 2015 workshop (\url{http://astrohackweek.org/2015/}), 
since the direction and the scope of this investigation was ---to some degree--- initiated 
through discussions in this workshop. Throughout this investigation, we have made use of 
publicly available software packages $\mathtt{emcee}$ and $\mathtt{abcpmc}$. We have also used the publicly available python implementation of the FoF algorithm $\mathtt{pyfof}$ (\url{https://github.com/simongibbons/pyfof}).



\bibliographystyle{mnras}
\bibliography{ccppabc}

\label{lastpage}
\end{document}